\documentclass[12pt]{article}
\usepackage{authblk}
\usepackage{here}
\usepackage{amsmath,amssymb}
\usepackage{bm}
\usepackage{graphicx}           % modified by SS
\usepackage{ascmac}
\usepackage{color}
\usepackage{braket}
\usepackage{here}
\usepackage{cite}
\usepackage[top=25truemm,bottom=30truemm,left=20truemm,right=20truemm]{geometry}% modified by SS
\usepackage{ytableau}
\usepackage{multirow}
\usepackage{afterpage}
\usepackage{cases}
\usepackage{fancybox}
\usepackage[driverfallback=dvipdfmx]{hyperref}
\usepackage{simplewick}
\hypersetup{colorlinks=true, citecolor=blue, linkcolor=blue, linkbordercolor=0 0 1}
\makeatletter

\@addtoreset{equation}{section}
\makeatother
\newlength{\myh}
\baselineskip=\normalbaselineskip
\newcommand{\cO}{{\mathcal O}}
\newcommand{\bC}{{\mathbb C}}
\newcommand{\bR}{{\mathbb R}}
\newcommand{\bS}{{\mathbb S}}
\newcommand{\bZ}{{\mathbb Z}}

\newcommand{\cN}{{\mathcal N}}

\newcommand{\wt}{\widetilde}

\newcommand{\nn}{\nonumber}

\newcommand{\Mkk}{M_{\rm KK}}

\newcommand{\ol}{\overline}  %overline
\newcommand{\tr}{\mathrm{tr}}  %tr %modified by SS
\newcommand{\NS}{\mathrm{NS}}  %NS
\newcommand{\dfbv}{\mathrm{D4_{BV}}}  %D4BV
\newcommand{\bv}{\mathrm{D5_{BV}}}  %D5BV
\newcommand{\ytil}{\tilde{y}}  %y tilde
\newcommand{\ad}{\dot{a}}  %a dot
\newcommand{\sq}[1]{\left[ #1 \right] }  %[...]
\newcommand{\som}{SO(4)_{123 z}}  %SO(4)1~3z
\newcommand{\soi}{SO(4)_{6789}}  %SO(4)6~9
\newcommand{\soyi}{SO(5)_{\ytil 6789}}  %SO(5)y6~9
\newcommand{\ds}{\oplus}  %direct sum
\newcommand{\soyianti}{\ytableausetup{smalltableaux} \ydiagram{1,1}_{10}}  %young diagram SO(5) rank 2 antisym
\newcommand{\soyivec}{\ytableausetup{smalltableaux} \ydiagram{1}_{5}}  %young diagram SO(5) vector
\newcommand{\soisym}{\ytableausetup{smalltableaux} \ydiagram{2}_{9}}  %young diagram SO(4) rank 2 sym
\newcommand{\soianti}{\ytableausetup{smalltableaux} \ydiagram{1,1}_{6}}  %young diagram SO(4) rank 2 antisym
\newcommand{\soivec}{\ytableausetup{smalltableaux} \ydiagram{1}_{4}}  %young diagram SO(4) vector
\newcommand{\singlet}{\textbf{1}}  %1
\newcommand{\half}{\frac{1}{2}}
\newcommand{\del}{\partial}
\newcommand{\deldel}[2]{\frac{\del #1}{\del #2}}
\newcommand{\drho}{\delta\rho}

\def\mat#1{\matt[#1]}
\def\matt[#1,#2,#3,#4]{\left(%
\begin{array}{cc} #1 & #2 \\ #3 & #4 \end{array} \right)} 
\begin{document}
\title{{\LARGE Stringy excited baryons in holographic QCD}}
%\author{{\Large Y. Hayashi, T.Ogino, T.Sakai}}
\author[1]{Yasuhiro Hayashi}
\author[1]{Takahiro Ogino}
\author[1,2]{Tadakatsu Sakai}
\author[3,4]{Shigeki Sugimoto}
\affil[1]{Department of Physics, Nagoya University, Nagoya 464-8602, Japan}
\affil[2]{Kobayashi-Maskawa Institute for the Origin of Particles 
and the Universe, Nagoya University, Nagoya 464-8602, Japan}
\affil[3]{Center for Gravitational Physics,
Yukawa Institute of Theoretical Physics, Kyoto University,
Kyoto 606-8502, JAPAN}
\affil[4]{Kavli Institute for the Physics and Mathematics
 of the Universe (WPI),
The University of Tokyo, Kashiwanoha, Kashiwa 277-8583, JAPAN}

\date{\today}
\maketitle
%
%\tableofcontents
%
\thispagestyle{empty}
\setcounter{page}{0}

\begin{abstract}
We analyze excited baryon states using a holographic dual of QCD
that is defined on the basis of an intersecting D4/D8-brane system.
Studies of baryons in this model have been made by regarding them as a 
topological soliton of a gauge theory on a five-dimensional
curved spacetime. However, this allows one to obtain only a certain
class of baryons. We attempt to present a framework such that
a whole set of excited baryons can be treated in a systematic
way. This is achieved by employing the original idea of Witten,
which states that a baryon is described by a system 
composed of $N_c$ open strings emanating from a baryon vertex.
We argue that this system can be formulated by an ADHM-type matrix model
of Hashimoto-Iizuka-Yi together with an infinite tower of the
open string massive modes.
Using this setup, we work out the spectra of excited baryons
and compare them with the experimental data.
In particular, we derive a formula for the nucleon Regge trajectory
assuming that the excited nucleons lying on the trajectory are 
characterized by the excitation of a single open string attached
on the baryon vertex.
\end{abstract}

\setcounter{section}{+0}
\setcounter{subsection}{+0}

\newpage

%\afterpage{\clearpage}
%\newpage
%
%%%%%%%%%%%%%%%%%%%%%%%%%%%%%%%%%%%%%%%%%%%%%%%%%%%%%%%%%%%%%%%%%%%
%
\section{Introduction}

Ever since the AdS/CFT correspondence was proposed by Maldacena
(for a review, see \cite{adscft}), it has been recognized
that it may provide us with a powerful tool for analyzing
nonperturbative dynamics of non-Abelian gauge theories.
One of the most intensive applications of the AdS/CFT correspondence
is to the hadron physics of QCD.
A key ingredient of hadron physics is how to understand 
spontaneous breaking of chiral symmetry.
A holographic dual of QCD (in the top down approach) with manifest
chiral symmetry was presented in \cite{SS1,SS2} on the basis of an
intersecting D4/D8-brane configuration. It was argued there that chiral
symmetry breaking is realized as a smooth interpolation of
D8 - anti-D8-brane ($\ol{\rm D8}$) pairs in a curved background
corresponding to D4-branes in type IIA supergravity.
The associated Nambu-Goldstone mode (pion) is shown to arise from
the 5 dimensional gauge field on the interpolated D8-branes.
This model is formulated in the large $N_c$ and large 't Hooft coupling
$\lambda$ regime with $N_c\gg N_f$,
where $N_c$ and $N_f$ are the numbers of colors and flavors, respectively,
for the purpose of suppressing intricate stringy and quantum gravity effects.
In spite of this approximation, the predictions of this model match well
with various experimental data in low-energy hadron physics.

In particular, it has been shown that the meson effective
theory is given by a 5 dimensional $U(N_f)$ gauge theory,
and a tower of vector and axial-vector mesons including
$\rho$ and $a_1$ mesons appear as the Kaluza-Klein (KK) modes
of the 5 dimensional gauge field. Other mesons including
higher-spin mesons are interpreted as excited open string modes
attached on the D8-branes. \cite{ISS} As they are described
by an open string, nearly linear Regge trajectories
with mild nonlinear corrections are obtained quite naturally,
and it has been argued that the predicted meson spectrum
agrees at least qualitatively with what is observed in nature.

The holographic model is also used to study the baryon sector.
This is performed by noting that a baryon can be realized as 
a topological soliton in the 5 dimensional gauge theory
with a baryon number identified with a topological number.
The original idea was due to Skyrme \cite{Skyrme} who
claimed that the baryons are solitons in a model given by adding
a so-called Skyrme term to the chiral Lagrangian of the massless pion.
In the holographic model, the soliton solution is given by
an instanton solution with the instanton number regarded as
the baryon number \cite{SS1}. Analysis of the moduli space
quantum mechanics analogous to the work \cite{ANW} in the Skyrme model
was performed in \cite{HSSY} and \cite{HSS} to obtain the baryon
spectrum and the static properties, respectively,\footnote{See also
\cite{Hong:2007kx,HMY,Kim:2008pw}.}
and again many of the results turned out to be consistent with
the experimental data.
However, one of the limitations in \cite{HSSY} is that
it describes only a subclass of baryons with $I=J$ for $N_f=2$.
Here, $J$ and $I$ denote the spin and the isospin of a baryon, 
respectively.
The reason for this limitation is clear:
the moduli space approximation only takes into account the light
degrees of freedom that correspond to the massless sector
in the open string spectrum.
We are led naturally to expect that incorporation of massive open
string states enables us to obtain a larger class of baryons with
$I\ne J$,\footnote{For another approach to holographic baryons with
$I\ne J$, see \cite{HMM}, which is based on the study
of a matrix model formulated in \cite{HIY}.}
as was done in \cite{ISS} for the meson sector.

The purpose of this paper is to examine holographic baryons following
this line.
To this end, we utilize the idea of Witten \cite{Witten:bv}
that a holographic description of baryons is made by introducing
a D-brane configuration, called a baryon vertex.
In the present holographic model, we add a D4-brane
that wraps around an $\bS^4$ with $N_c$ units of RR-flux over it. 
It was found in \cite{Witten:bv,Gross:1998gk} that the RR-flux forces
$N_c$ open strings to extend between the D4-brane and the D8-branes.
The whole system is regarded as a holographic baryon.
As a consistency check, the instanton
solution is identical to the baryon vertex D4-brane
in the context of the effective theory.
The baryon states can be computed by working out
a bound state of a many-body quantum mechanics that is defined
from open strings attached on the baryon vertex.
There are two types of open strings that should be
taken into account. One of them is the 4-4 strings
with both end points attached on the baryon vertex
D4-brane and the other is the 4-8 strings that extend
between the D4-brane and one of the D8-branes.
As shown in \cite{Witten:1995gx,Douglas:1995bn},
the massless degrees
of freedom that arise from these strings correspond to
the instanton moduli space in the ADHM construction \cite{Atiyah:1978ri}
and it is expected to be equivalent to the moduli space quantum
mechanics in the soliton approach. This approach was proposed in
\cite{HIY}, in which a matrix quantum mechanics describing
multiple baryon systems was derived. Our main idea is to incorporate
the massive open string states into this quantum mechanics
to describe heavier baryons.
Solving the bound state problem in quantum mechanics is highly
involved in general. In this present case, however, we argue that
taking the large $N_c$ limit makes the problem
tractable. This is because the string coupling is
of $\cO(1/N_c)$ so that interactions among open strings
are mostly negligible in the large $N_c$ limit.

The fundamental degrees of freedom in the quantum mechanics
are given by massless and an infinite tower of massive modes
of open strings attached on the baryon vertex D4-brane.
The mass spectrum can be worked out by quantizing the open strings
in the curved background (\ref{metric}), but this is 
technically difficult to achieve. As suggested in \cite{ISS},
this problem gets simplified
drastically by taking the limit $\lambda\gg 1$, where the spacetime
curvature becomes negligible.
Nontrivial curvature effects in the mass spectrum are incorporated
perturbatively in $1/\lambda$ expansions.
Using these results, the many-body quantum mechanics is formulated 
in a manner that is simple and powerful enough to study 
a wide range of holographic baryons quantitatively.
As an application, we derive the mass formula of the nucleon and its
excited states. We also discuss its implication to the 
nucleon Regge trajectory.

The organization of this paper is as follows.
In section 2, after giving a brief review of the holographic model
of QCD with the emphasis on a baryon vertex, 
we compute the mass spectrum of the open strings attached
on the baryon vertex and D8-branes.
With this result, section 3 formulates a many-body quantum mechanics
that enables one to compute the mass spectrum of baryons
that are missing in \cite{HSSY}.
In section 4, we compare the predictions of this model
to experiments. We conclude in section 5 with a summary
and some comments about future directions.
Some technical formulas that are used in the paper are
summarized in appendix A.

%%%%%%%%%%%%%%%%%%%%%%%%%%%%%%%%%%%%%%%%%%%%%%%%%%%%%%%%%%%%%%%%%%%
%
\section{Holographic model of QCD and baryons}
\label{sec:baryonstates}

\subsection{Brief review of the model}

The holographic model of QCD we work with is constructed
from an intersecting D4/D8-brane system \cite{SS1,SS2}.
The $N_c$ D4-branes wrap around a circle on which a SUSY-breaking
boundary condition is imposed, and yield gluons of gauge group $SU(N_c)$
on the worldsheet at low energy compared with the circle
radius $1/\Mkk$.
$N_f$ D8- and $\ol{\rm D8}$-branes are placed at the anti-podal
points of the SUSY breaking circle.
Quantization of D4-D8 and D4-$\ol{\rm D8}$ strings gives 
left- and right-handed quarks in the fundamental representation
of $SU(N_c)$, respectively
This system has a manifest chiral $U(N_f)_L\times U(N_f)_R$
symmetry.

The holographic dual of this model is formulated by replacing
the D4-branes with a solution of type IIA supergravity
with a nontrivial dilaton $\phi$ \cite{Witten:d4}:
\begin{align}
&ds^2=\frac{4}{27}\lambda l_s^2 \,d\wt{s}^2\ ,
\nn\\
&d\wt{s}^2=
K(r)^{1/2} \eta_{\mu\nu}dx^{\mu}
dx^{\nu}+ K(r)^{-5/6}dr^2 + K(r)^{-1/2}r^2d\theta^2
+\frac{9}{4}K(r)^{1/6}d\Omega_4^2\ ,
\label{metric}
\end{align}
\begin{eqnarray}
 e^\phi= \frac{\lambda^{3/2}}{3\sqrt{3}\pi N_c}K(r)^{1/4}\ .
\label{dilaton}
\end{eqnarray}
Here, $\mu,\nu=0,1,2,3$ denote the indices of the 4 dimensional
Minkowski spacetime where QCD is defined, $d\Omega_4^2$ is the metric of
a unit $\bS^4$, and $K=1+r^2$.\footnote{
The radial coordinate $r$ is related to $U/U_{\rm KK}$ used in
\cite{SS1,SS2}
by $(U/U_{\rm KK})^3=1+r^2$.}
$\theta$ is the coordinate of the SUSY-breaking circle.
In addition, there exist $N_c$ units RR-4-form flux over the
$\bS^4$:
\begin{align}
\frac{1}{2\pi}\int_{S^4}F_4=N_c\ .
\label{RR}
\end{align}
It is useful to define
\begin{align}
  z=r\sin\theta \ ,~~  y=r\cos\theta \ .\nn
\end{align}
The metric (\ref{metric}) is defined in the decoupling limit, where
the dependence on $l_s$, the string length, factorizes as a prefactor.
As a consequence, the string theory on this background is independent
of $l_s$. This allows one to set
\begin{eqnarray}
\alpha'\equiv l_s^2=\frac{27}{4\lambda}
\label{lslam}
\end{eqnarray}
in units of $\Mkk=1$ so that $ds^2=d\wt s^2$. (See \cite{ISS} for more
details on this point.)
It follows that the stringy excitation modes have mass of 
$\cO(\lambda^{1/2})$ and may be neglected
at low energies for $\lambda\gg 1$.

Assuming $N_c \gg N_f$,\footnote{For this, we mean that
we consider $N_f$ to be of $\cO(1)$ and only take into account
the leading terms in the $1/N_c$ expansion.
}
the D8-branes can be regarded as
probes with no backreaction to the metric (\ref{metric}) taken
into account. It has been shown \cite{SS1} that the D8- and 
$\ol{\rm D8}$-brane pairs interpolate with each other smoothly at 
$z=y=0$ and the resultant D8-brane worldvolume is specified by 
the embedding equation $y=0$.
In this setup, the mesons are identified with the open strings
attached on the D8-branes that can move along the $z$ direction.

In order to incorporate baryon degrees of freedom into the model, 
we introduce a baryon vertex \cite{Witten:bv}, which
is given by a single D4-brane wrapping around $\bS^4$
at $z=y=0$.
We refer to this D4-brane as a D4$_{\rm BV}$ in order to
distinguish it from $N_c$ color D4-branes.
The RR flux (\ref{RR}) forces $N_c$ open strings to extend
between the D8-branes and the baryon vertex. This configuration
is identified with a single baryon.
It is argued in \cite{SS1} that this brane system is 
realized as an instanton solution on the
D8-brane worldvolume theory. By analyzing the moduli
space quantum mechanics corresponding to this
instanton solution, Refs. \cite{HSSY,HSS} showed
that aspects of the baryon dynamics are reproduced from
this model both qualitatively and quantitatively.
One of the limitations in this analysis, however, is that
describing a baryon vertex as a classical solution
of the $U(N_f)$ gauge theory on the D8-branes is valid only 
for low-lying baryons. This is because the $U(N_f)$ gauge theory
is an effective theory of the D8-branes with only the massless
degrees of freedom taken into account. In addition,
the moduli space approximation only keeps light degrees
of freedom in the fluctuations around the soliton solution.
In fact, these are the main reasons why the analysis in \cite{HSSY}
leads to only baryons with the spin $J$ and isospin $I$ equal to
each other for the $N_f=2$ case.
{}For the purpose of obtaining more general baryons, we thus have
to consider stringy effects in the baryon vertex.

\subsection{Quantization of open strings in a flat spacetime limit}
\label{subsec:states}

It is highly difficult to make a full quantization of a string that 
propagates in the curved background (\ref{metric})
in the presence of the RR flux (\ref{RR}).
In order to circumvent this problem, we follow \cite{ISS}.
We first take the large $\lambda$ limit, where
the curved background can be approximated with a 10 dimensional
flat spacetime.
Then, the baryon configuration reduces to a system with
$N_f$ D8-branes and a D4$_{\rm BV}$-brane with $N_c$ open strings
stretched between them in the flat background.
{}For a technical reason, it is useful to formally T-dualize the system
in the $y$ direction. The D8/D4$_{\rm BV}$-brane system gets mapped to
the D9/D5$_{\rm BV}$-brane 
configuration shown below:
\newlength{\lnficonf}
\setlength{\lnficonf}{0.7cm}
\begin{table}[H]
\begin{center}
  \begin{tabular}{|c|c|c|c|c|c|c|c|c|c|c|} \cline{1-11}
	\parbox[c][\lnficonf][c]{0cm}{}
	 & 0 & 1 & 2 & 3 & $z$ & $\ytil$ & 6 & 7 & 8 & 9 \\ \cline{1-11}
	\parbox[c][\lnficonf][c]{0cm}{}    
 	 $N_f\times$D9 & $\bigcirc$  & $\bigcirc$ & $\bigcirc$ & $\bigcirc$ & $\bigcirc$ & $\bigcirc$ & $\bigcirc$ & $\bigcirc$ & $\bigcirc$ & $\bigcirc$ \\ \cline{1-11}
	\parbox[c][\lnficonf][c]{0cm}{}    
	 D5$_{\rm BV}$ & $\bigcirc$  &  &  &  &  & $\bigcirc$ & $\bigcirc$ & $\bigcirc$ & $\bigcirc$ & $\bigcirc$ \\ \cline{1-11}
  \end{tabular}
\parbox{15cm}{
 \caption{D9/D5$_{\rm BV}$-brane system. $\ytil$ is the T-dualized
 coordinate of $y$.}
\label{table:95conf}}
\end{center}
% \vspace{-0.5cm}
\end{table}
The $123z$- and $6789$-directions are labeled by indices $M$ 
and $i$, respectively.
The $6789$-directions span $\bR^4$, which results from the $\bS^4$
that is decompactified for $\lambda\gg 1$.
Quantization of a 9-5 and 5-5 string is performed most easily
by using a light-cone quantization, where the light-cone
coordinate is taken to be $x^0\pm \ytil$.
The manifest spacetime symmetry of the brane system
is $SO(4)_{123z}\times SO(4)_{6789}$.

We first study the light-cone quantization of a 9-5 string.
The equations of motion (EOM)
of the worldsheet boson in the 6789-directions is solved
in terms of Fourier expansions with an integer modding, while that in the
123$z$-directions in terms of those with a half-integer modding,
because of the boundary conditions imposed on them.
{}For the worldsheet fermions in the NS (R) sector,
the solutions of the EOM in the 6789-directions are written in terms of
Fourier expansions with a half-integer (integer) modding, while
those in the 123$z$-directions written in terms of
those with an integer (half-integer) modding.
It follows that the NS ground state is degenerate due to the fermion
zero modes, belonging to a spinor representation of $SO(4)_{123z}$.
The R ground state is degenerate too, and
belongs to a spinor representation of $SO(4)_{6789}$.
We label an irreducible representation of $SO(4)_{123z}\simeq
(SU(2)_L\times SU(2)_R)/\bZ_2$ by $(s_L,s_R)$,
where $s_L$ and $s_R$ are the spin of $SU(2)_L$ and $SU(2)_R$,
respectively. The (integer spin) representation
of $SO(4)_{6789}$ is labeled by Young tableaux as
$\singlet$, $\soivec$, $\soianti$, $\soisym$, etc.,where
the subscripts denote the dimensions.
Then, the low-lying 9-5 string states in the NS sector 
with the GSO projection imposed
are summarized in Table \ref{table:single}.
\begin{table}[h]
%\vspace{0.1cm}
\begin{center}
  \begin{tabular}{|c|c|c|c|}
	\multicolumn{2}{c}{} 
	& \multicolumn{1}{c}{$\som$} 
	& \multicolumn{1}{c}{$\soi$} \\ \hline
	\parbox[c][1cm][c]{0cm}{} $N_{95}=0$
	& $\ket{a}_{\NS}$ 
	& $(1/2,0)$ 
	& $\singlet$ \\ \hline
	\parbox[c][1cm][c]{0cm}{} $N_{95}=1/2$ 
	& $\alpha_{-1/2}^M \ket{a}_{\NS}$ 
	& $(0,1/2) \ds (1,1/2)$ 
	& $\singlet$ \\
	\parbox[c][1cm][c]{0cm}{} 
	& $\psi_{-1/2}^i \ket{\ad}_{\NS}$ 
	& $(0,1/2)$ 
	& $\soivec$ \\ \hline
	\parbox[c][1cm][c]{0cm}{} $N_{95}=1$ 
	& $\alpha_{-1}^i \ket{a}_{\NS}$ 
	& $(1/2,0)$ 
	& $\soivec$ \\
	\parbox[c][1cm][c]{0cm}{} 
	& $\psi_{-1}^M \ket{\ad}_{\NS}$ 
	& $(1/2,0) \ds (1/2,1)$ 
	& $\singlet$ \\
	\parbox[c][1cm][c]{0cm}{} 
	& $\alpha_{-1/2}^M \alpha_{-1/2}^N \ket{a}_{\NS}$ 
	& $(1/2,0) \ds (1/2,1) \ds (3/2,1)$ 
	& $\singlet$ \\
	\parbox[c][1cm][c]{0cm}{} 
	& $\alpha_{-1/2}^M \psi_{-1/2}^i \ket{\ad}_{\NS}$ 
	& $(1/2,0) \ds (1/2,1)$ 
	& $\soivec$ \\
	\parbox[c][1cm][c]{0cm}{} 
	& $\psi_{-1/2}^i \psi_{-1/2}^j \ket{a}_{\NS}$ 
	& $(1/2,0)$ 
	& $\raisebox{0.1cm}{$\soianti$}$ \\ \hline
  \end{tabular}
\parbox{15cm}{
\caption{Low-lying 9-5 string states in the NS sector. $\alpha_{-r}$ denotes
the Fourier mode of a worldsheet scalar and $\psi_{-r}$ that of a worldsheet
fermion. $M=1,2,3,z$ and $i=6,7,8,9$ are the vector indices for
$SO(4)_{123z}$ and $SO(4)_{6789}$, respectively.
$a$ and $\dot{a}$ are the undotted and dotted spinor indices of
$SO(4)_{123z}\simeq (SU(2)_L\times SU(2)_R)/\bZ_2$, corresponding to the 
doublet representation of $SU(2)_L$ and $SU(2)_R$, respectively.
$N_{95}$ is the total excitation number of a 9-5 string state
with the mass squared equal to $N_{95}/l_s^2$. }
\label{table:single}}
\end{center}
\end{table}
Although it is not manifest in the light-cone quantization,
the 6 dimensional Lorentz symmetry on the $\bv$-brane worldvolume
allows one to summarize the massive excitations into the
irreducible representations of the little group $SO(5)_{\ytil 6789}$,
which contains $\soi$ as a subgroup.
Table \ref{table:soyi} gives a list of the low-lying 9-5 string states
in the NS sector in terms of $SO(5)_{\ytil 6789}$.
\begin{table}[h]
%\vspace{0.1cm}
\begin{center}
  \begin{tabular}{|c|c|}
 \multicolumn{1}{c}{}
	& \multicolumn{1}{c}{$\som\times\soyi$} \\ \hline
	\parbox[c][1cm][c]{0cm}{} 
	$N_{95}=0$
	& $(1/2,0)\,\singlet$ \\ \hline
	\parbox[c][1cm][c]{0cm}{}
	$N_{95}=1/2$
	& $(0,1/2)\,\soyivec \ds (1,1/2)\,\singlet $ \\ \hline
	\parbox[c][1cm][c]{0cm}{} 
	\multirow{2}{*}{$N_{95}=1$}  
	& $(1/2,0)\,\soyianti \ds (1/2,0)\,\soyivec \ds (1/2,1)\,\soyivec$ \\
	\parbox[c][1cm][c]{0cm}{} 
	& $\ds\, (1/2,0)\,\singlet \ds (1/2,1)\,\singlet \ds (3/2,1)\,\singlet$ \\ \hline
  \end{tabular}
\parbox{15cm}{
 \caption{Low-lying 9-5 string states in the NS sector (states
 in Table \ref{table:single}) classified by
 $SO(4)_{123z}\times SO(5)_{\ytil 6789}$.}
 \label{table:soyi}}
\end{center}
\end{table}

We next study the mass spectrum of a 5-5 string using the light-cone
quantization.
The worldsheet bosons can be Fourier expanded with 
an integer modding for both $123z$- and $6789$-directions.
The worldsheet fermions in the NS (R) sector
can be Fourier expanded with a half-integer (integer) modding
for the $123z$- and $6789$-directions.
The physical ground state in the NS sector is massless and
given by 
$\psi_{-1/2}^M|0\rangle_{\rm NS}$ and $\psi_{-1/2}^i|0\rangle_{\rm NS}$.
Here, $|0\rangle_{\rm NS}$ is tachyonic, being GSO-projected out.
The first excited 5-5 string states in the NS sector 
that survive the GSO projection are given by acting on $|0\rangle_{\rm NS}$
with a set of the creation operators with
total excitation number equal to $3/2$.
These have the mass squared $(3/2-1/2)/l_s^2=1/l_s^2$
and are listed in Table \ref{table:55ns}.
\begin{table}[h]
%\vspace{0.1cm}
\begin{center}
  \begin{tabular}{|c|c|c|c|}
	\multicolumn{2}{c}{} 
	& \multicolumn{1}{c}{$\som$} 
	& \multicolumn{1}{c}{$\soi$} \\ \hline
	\parbox[c][1cm][c]{0cm}{} $N_{55}=0$ 
	& $\psi_{-1/2}^M \ket{0}_{\NS}$ 
	& $(1/2,1/2)$
	& $\singlet$ \\
	\parbox[c][1cm][c]{0cm}{} 
	& $\psi_{-1/2}^i \ket{0}_{\NS}$ 
	& $(0,0)$ 
	& $\soivec$ \\ \hline
	\parbox[c][1cm][c]{0cm}{} $N_{55}=1$ 
	& $\psi_{-1/2}^M\psi_{-1/2}^N\psi_{-1/2}^L \ket{0}_{\NS}$ 
	& $(1/2,1/2)$ 
	& $\singlet$ \\
	\parbox[c][1cm][c]{0cm}{} 
	& $\psi_{-1/2}^M\psi_{-1/2}^N\psi_{-1/2}^i \ket{0}_{\NS}$ 
	& $(1,0)\ds(0,1)$ 
	& $\soivec$ \\
	\parbox[c][1cm][c]{0cm}{} 
	& $\psi_{-1/2}^M\psi_{-1/2}^i\psi_{-1/2}^j \ket{0}_{\NS}$ 
	& $(1/2,1/2)$ 
	& $\soianti$ \\
	\parbox[c][1cm][c]{0cm}{} 
	& $\psi_{-1/2}^i\psi_{-1/2}^j\psi_{-1/2}^k \ket{0}_{\NS}$ 
	& $(0,0)$ 
	& $\ytableausetup{smalltableaux} \ydiagram{1,1,1}_{4}$ \\
	\parbox[c][1cm][c]{0cm}{} 
	& $\psi_{-3/2}^M \ket{0}_{\NS}$ 
	& $(1/2,1/2)$ 
	& $\singlet$ \\
	\parbox[c][1cm][c]{0cm}{} 
	& $\psi_{-3/2}^i \ket{0}_{\NS}$ 
	& $(0,0)$ 
	& $\soivec$ \\
	\parbox[c][1cm][c]{0cm}{} 
	& $\alpha_{-1}^M \psi_{-1/2}^N \ket{0}_{\NS}$ 
	& $(0,0) \ds (1,0) \ds (0,1) \ds (1,1)$ 
	& $\singlet$ \\
	\parbox[c][1cm][c]{0cm}{} 
	& $\alpha_{-1}^M \psi_{-1/2}^i \ket{0}_{\NS}$ 
	& $(1/2,1/2)$ 
	& $\soivec$ \\
	\parbox[c][1cm][c]{0cm}{} 
	& $\alpha_{-1}^i \psi_{-1/2}^M \ket{0}_{\NS}$ 
	& $(1/2,1/2)$ 
	& $\soivec$ \\
	\parbox[c][1cm][c]{0cm}{} 
	& $\alpha_{-1}^i \psi_{-1/2}^j \ket{0}_{\NS}$ 
	& $(0,0)$ 
	& $\soivec \ds\raisebox{0.1cm}{$\soianti$} \ds \soisym$ \\ \hline
  \end{tabular}
\parbox{15cm}{
\caption{Low-lying 5-5 string states in the NS sector.
$N_{55}$ is the total excitation number of a 5-5 string state
with the mass squared equal to $N_{55}/l_s^2$. 
}\label{table:55ns}}
\end{center}
\end{table}

As in the 9-5 string states, any massive state of the 5-5 string is
summarized into an irreducible representation of 
$SO(4)_{123z}\times SO(5)_{\ytil 6789}$. It is found that the 
first excited states with $N_{55}=1$ in Table \ref{table:55ns}
are rearranged as
\begin{align}
\sq{(1,0) \ds (0,1)}&\,\ytableausetup{smalltableaux} \ydiagram{1}_{5}\ ,\ \ \ 
(1/2,1/2)\,\ytableausetup{smalltableaux} \ydiagram{1,1}_{10},\ \ \ 
{(1/2,1/2)\,\ytableausetup{smalltableaux} \ydiagram{1}_{5}}\ ,\ \ \ 
(0,0)\,\ytableausetup{smalltableaux} \ydiagram{1,1,1}_{10}\ ,\ \ \  
\nn\\
(0,0)&\,\ytableausetup{smalltableaux} \ydiagram{2}_{15}\ , \ \ 
{(1/2,1/2)\,\singlet}\ ,\ \ \ 
\sq{(0,0)\ds(1,1)}\,\singlet \ , \ \ \ 
\label{55:GSO1st}
\end{align}
where the Young tableaux are those of $SO(5)_{\ytil 6789}$.
In fact, these states are obtained as the decomposition
of $~\ytableausetup{smalltableaux} \ydiagram{2}_{44} \ds~
\ytableausetup{smalltableaux} \ydiagram{1,1,1}_{84}$ of
$SO(9)$, which is the same as the first excited 9-9 string states
considered in \cite{ISS}.

%
%******************************************************************
%
\subsection{Symmetries in the presence of a baryon vertex}
\label{subsec:disc}

Reference \cite{ISS} discusses that the D4/D8-brane system
has discrete symmetries that are identified with those in 
massless QCD. The parity $P$ and charge conjugation $C$ are given by
\begin{eqnarray}
P=I_{123z}\ ,~~~ C=I_{z89}\Omega\,(-1)^{F_L}\ ,
\end{eqnarray}
respectively, where $I_{i_1i_2\cdots}$ is spacetime involution along the
$i_1,i_2,\cdots$ directions,
$\Omega$ is a worldsheet parity, and $F_L$ is a spacetime
fermion number in the left-moving sector of a string worldsheet.
A $\dfbv$-brane placed at $x^1=x^2=x^3=y=z=0$\footnote{
For a $\dfbv$-brane wrapped on $\bS^4$, it can be shown that
$y=z=0$ is energetically favored and realized in the
classical minimal energy configuration. It may be located
anywhere in $\bR^3\ni x^{1,2,3}$, because of the translational
invariance. Here we just put it at $x^1=x^2=x^3=0$ to have
a $P$-invariant configuration.}
is invariant under $P$, while it is mapped to a
$\ol{\rm D4}_{\rm BV}$-brane under $C$.
To see the latter, note that when the $\bZ_2$ action
generated by $C$ is gauged, a background has an O6-plane at
$z=x^8=x^9=0$, and it is known that the $\dfbv$-brane has to be paired
with a $\ol{\rm D4}_{\rm BV}$-brane in the presence of the
O6-plane.\cite{ISS2}
This is consistent with the fact that the baryon is invariant under
the parity, up to sign of the wavefunction, while it
is mapped to an anti-baryon under the charge conjugation.

In order to see how $P$ acts on the NS ground state of the
9-5 string considered in section \ref{subsec:states},
it is useful to write the parity operator in a bosonized form.
We note that the worldsheet fermions
of a 9-5 string can be expressed using free worldsheet
complex scalars $H^1$ and $H^2$ as
\begin{align}
\psi^1 \pm i \psi^2 =e^{\pm i H^1} \ ,~~
\psi^3 \pm i \psi^z =e^{\pm i H^2} \ .\nn
%\label{eq:comp}
\end{align}
Parity acts on the worldsheet fermions as
\begin{align}
\psi^M \to - \psi^M \ ,\nn
%\label{eq:Ppsi}
\end{align}
which in turn induces the transformation of
$H^1,H^2$ as
\begin{align}
(H^1,H^2) \to (H^1 + (2n_1+1)\pi, H^2 + (2n_2+1)\pi)\ ,
\label{eq:PHp}
\end{align}
with a choice of $n_1,n_2\in\bZ$.
The vertex operator corresponding to the
NS ground state of a 9-5 string is given by
\begin{align}
e^{i(s_1H^1+s_2H^2)} \ ,
\end{align}
up to a ghost sector that is invariant under $P$,
with $s_1=s_2=\pm 1/2$ for $\ket{a}_{\NS}$
and $s_1=-s_2=\pm 1/2$ for $\ket{\dot a}_{\NS}$.
Therefore, the parity transformation (\ref{eq:PHp}) acts
as the chirality operator on the spinor representation
of $SO(4)_{123z}$ up to a sign ambiguity.
We choose $n_1$ and $n_2$ in (\ref{eq:PHp}) such that
$\ket{a}_{\NS}$ and $\ket{\dot a}_{\NS}$ are
parity even and odd, respectively. With this
convention, the parity of the proton and
the neutron turn out to be even. This is consistent
with the conventional choice of the parity in QCD,
in which the parity of quarks are chosen to be even.
For a $\ol{\rm D5}_{\rm BV}$-brane, which represents
an anti-baryon, since the GSO projection is opposite,
the parity of the the NS ground state is odd.
This is again consistent with the fact that the anti-quarks
have odd parity.

Then, the parity of the excited states can be computed
by using the transformation laws of the creation operators
that act on the ground state. Namely, $\psi_{-r}^M$ and $\alpha_{-r}^M$
with $M=1,2,3,z$ are parity odd and
$\psi_{-r}^i$ and $\alpha_{-r}^i$ with $i=6,7,8,9$ are parity even
operators.

In addition to these symmetries, the D4/D8-system admits a discrete
symmetry that has no counterpart in QCD.
This is called $\tau$-parity\footnote{$\tau$-parity was originally
introduced in \cite{BMT} in the context of glueball spectrum
and then generalized to the system with quarks in \cite{ISS}.
}
and defined as
\begin{align}
  P_\tau=I_{y9}\,(-1)^{F_L} \ .
\end{align}
As discussed in \cite{ISS}, 
both the quarks that originate from 4-8 and 4-$\bar{8}$ strings
in the open string picture and 
the gluons that originate from the 4-4 strings are even
under $\tau$-parity.
This implies that all the states that can be interpreted
as the genuine color singlet states of QCD have to be
$\tau$-parity even as well.
There are $\tau$-parity odd states in the spectrum of
the bound states in our model. However, such states
are artifacts of the model, which do not have counterparts in QCD,
and we will not consider them in the following.

Assuming that the $\dfbv$-brane is placed at $y=0$,
one can show that the $\dfbv$-brane is invariant
under the $\tau$-parity $P_\tau$.
To see this, we note that $I_{y9}$ maps
the D4$_{\rm BV}$ to a $\overline{\rm D4}_{\rm BV}$ 
and $(-1)^{F_L}$ maps it back to a D4$_{\rm BV}$.

{}For the purpose of reading off the $\tau$-parity of an open
string state, it is useful to work in the T-dualized description
used in section \ref{subsec:states}.
When the $y$-direction is T-dualized,
$P_\tau$ is mapped to
\begin{align}
  \widetilde{P}_\tau=I_{9\tilde{y}} \ ,
\label{eq:tauparity}
\end{align}
where $\tilde{y}$ is the T-dualized coordinate of $y$.
This is simply a $180^\circ$ rotation in the
9-$\ytil$ plane and it is easy to find the action of
$\wt P_\tau$ from the representation of $SO(5)_{\wt 6789}$
listed in Table \ref{table:soyi} and (\ref{55:GSO1st}).

In addition to the $\tau$-parity discussed above,
we can also use the $SO(5)$ isometry of $\bS^4$ in the background
to single out the open string states that could be used to
construct a baryon in QCD. It is easy to see that both
quarks and gluons are invariant under this $SO(5)$,
and hence the baryons in QCD have to be an $SO(5)$ singlet.
In the flat spacetime limit, the requirement
of the $SO(5)$ invariance amounts to
demanding that the states be $SO(4)_{6789}$-singlet and
carry no momentum along the 6789-directions.
In the T-dualized picture, we should also impose the condition that
the momentum along $\ytil$ is zero, since
the original $y$ direction is not compactified
and there is no winding mode along $y$.
Therefore, among the open string states obtained in
section \ref{subsec:states}, we only consider the
states that are invariant under $SO(4)_{6789}$ and
the $\tau$-parity $\wt{P}_\tau$,
and carry no momentum along the $\ytil 6789$ directions.

\subsection{Summary of the results}

We first derive the 9-5 string states that
meet the conditions discussed in the last subsection.
The requirement of $SO(4)_{6789}$ invariance implies
that the R-sector must be removed because all the states in the R-sector
are $SO(4)_{6789}$-nonsinglet.
It follows from the $\tau$-parity condition that 
among the $SO(4)_{6789}$-singlet NS states,
only those with an even number of the spacetime index 
$\tilde{y}$ are allowed. 
The NS ground state satisfies these conditions.
{}For the first excited states (those with $N_{95}=1/2$) 
listed in Table \ref{table:soyi}, only
the state with $(s_L,s_R)=(1,1/2)$ is allowed.
{}From the second excited states with $N_{95}=1$, we pick up
\begin{align}
(1/2,0)\singlet \ds (1/2,1)\singlet \ds (3/2,1)\singlet
\ . \nn
%\label{eq:secrem}
\end{align}
{}Finally, we set the momenta along the $\ytil 6789$ direction
to zero, which is equivalent to omitting the dependence
of the corresponding wavefunctions on $\ytil$ and $x^{6,7,8,9}$. 
These results are summarized in Table \ref{table:95summary},
%\newlength{\lrem}
%\setlength{\lrem}{1cm}
\begin{table}[h]
\begin{center}
  \begin{tabular}{|c|c|c|c|c|}
  \multicolumn{1}{c}{} &
  \multicolumn{1}{c}{$SO(4)_{123z}$} &
  \multicolumn{1}{c}{$SU(2)_J$} &
  \multicolumn{1}{c}{Parity} &
  \multicolumn{1}{c}{label $j$} 
\\ \hline
%&& patrity \\\hline
%	\parbox[c][\lrem][c]{0cm}{} 
	$N_{95}=0$
	& $(1/2,0)$
	& $1/2$
	& $+$  
	&  \\ \hline
%	\parbox[c][\lrem][c]{0cm}{}
	$N_{95}=1/2$ 
	& $(1,1/2)$
	& $3/2 \ds 1/2$
	& $-$  
	& $1$ \\ \hline
%	\parbox[c][\lrem][c]{0cm}{}
	$N_{95}=1$ 
	& $(1/2,0)$ 
	& $1/2$ 
	& $+$ 
	& $2$ \\
%	\parbox[c][\lrem][c]{0cm}{}
	& $(1/2,1)$
	& $3/2\ds 1/2$
	& $+$ 
	& $3$ \\
%	\parbox[c][\lrem][c]{0cm}{}
	& $(3/2,1)$ 
	& $5/2\ds 3/2 \ds 1/2$ 
	& $+$ 
	& $4$ \\ \hline
  \end{tabular}
\parbox{15cm}{
\caption{
9-5 string states that could contribute to genuine QCD
baryons. All the states belong to the fundamental representation of the
flavor $U(N_f)$ symmetry and have the unit charge with respect
to the $U(1)$ gauge symmetry on the $\dfbv$-brane. The massive 9-5 string 
states are labeled by $j=1,2,\cdots$, which will be used in section
\ref{Regge}.
}\label{table:95summary}}
\end{center}
\end{table}
where we also list the representation (spin) of $SU(2)_J$,
which is related to the $SO(3)_{123}$ subgroup of $SO(4)_{123z}$ by
$SU(2)_J/\bZ_2\simeq SO(3)_{123}$. Note that $SO(4)_{123z}$ symmetry
appears only in the flat spacetime limit and it is broken to
$SO(3)_{123}$ due to the $z$-dependence of the background.
The masses of these states in the flat spacetime limit are
proportional to the excitation number $N_{95}$ as
\begin{eqnarray}
 m^2=\frac{N_{95}}{\alpha'}=\frac{4\lambda}{27}N_{95}\ ,~~~
(N_{95}=0,1/2,1,\cdots )\ ,
\label{95mass}
\end{eqnarray}
where we have used the relation (\ref{lslam}).

The quantum field corresponding to the 9-5 massless state
is denoted by $\omega^I_\alpha$, which reduces to a function of time $t$ only
as discussed above. Here $\alpha=1,2$ is the spin index for $SU(2)_J$
and $I=1,2,\cdots,N_f$ is the index for the flavor $U(N_f)$ symmetry.

Next, we discuss the 5-5 string states.
As in the 9-5 string case, all the R-states are non-singlet 
under $SO(4)_{6789}$ and thus ruled out.
The NS massless states that satisfy all the conditions are given 
by $\psi_{-1/2}^M\ket{0}_{\NS}$ ($M=1,2,3,z$) only. 
The corresponding fields are denoted as $X^M$.
Again, these fields reduce to functions of $t$.
Among the first excited states with $N_{55}=1$
listed in (\ref{55:GSO1st}), 
the following states satisfy all the conditions
\begin{align}
2\,(0,0) \ds (1/2,1/2) \ds (1,1) \ .
\end{align}
%\newlength{\lrem}
%\setlength{\lrem}{1cm}
\begin{table}[h]
\begin{center}
  \begin{tabular}{|c|c|c|c|c|}
  \multicolumn{1}{c}{} &
  \multicolumn{1}{c}{$SO(4)_{123z}$} &
  \multicolumn{1}{c}{$SU(2)_J$} &
  \multicolumn{1}{c}{Parity} &
  \multicolumn{1}{c}{label $k$}
\\ \hline
%&& patrity \\\hline
%	\parbox[c][\lrem][c]{0cm}{} 
	$N_{55}=0$
	& $(1/2,1/2)$
	& $1\ds 0$
	& $-$ 
	&  \\ \hline
%	\parbox[c][\lrem][c]{0cm}{}
	$N_{55}=1$ 
	& $2\,(0,0)$
	& $0\ds 0$
	& $+$ 
	& $1,2$ \\
%	\parbox[c][\lrem][c]{0cm}{}
	& $(1/2,1/2)$ 
	& $1 \ds 0$ 
	& $-$ 
	& $3$ \\
%	\parbox[c][\lrem][c]{0cm}{}
	& $(1,1)$
	& $2 \ds 1 \ds 0$
	& $+$ 
	& $4$ \\ \hline
  \end{tabular}
\parbox{15cm}{
\caption{5-5 string states that could contribute to genuine QCD
baryons. All the states are singlet under the
flavor $U(N_f)$ symmetry and neutral under the $U(1)$ gauge symmetry
on the $\dfbv$-brane.
The massive 5-5 strings are labeled by $k=1,2,\cdots$, which will be
used in section \ref{Regge}.
}
\label{table:55summary}}
\end{center}
\end{table}
Note here that there are two $(0,0)$ states, and one of them comes
from $(0,0)\ytableausetup{smalltableaux}\,\ydiagram{2}_{15}$
in (\ref{55:GSO1st}) with two $\ytil$ indices.
The masses are given by
\begin{eqnarray}
 m^2=\frac{N_{55}}{\alpha'}=\frac{4\lambda}{27}N_{55}\ ,~~~
(N_{55}=0,1,2,\cdots )\ .
\label{55mass}
\end{eqnarray}
The results for the 5-5 strings are summarized in Table
\ref{table:55summary}.

%%%%%%%%%%%%%%%%%%%%%%%%%%%%%%%%%%%%%%%%%%%%%%%%%%%%%%%%%%%%%%%%%%%
%
\section{One-baryon quantum mechanics}
\label{sec:quant}

In the previous section, we obtained the spectrum of
the open strings attached on the baryon vertex
$\dfbv$-brane.\footnote{
In this and the following sections, we consider the original
D8/$\dfbv$ system, rather than the T-dualized version (D9/$\bv$ system)
considered in the previous section. Therefore, the 9-5 and 5-5
strings in the previous section correspond to 8-4 and 4-4 strings,
respectively.
}
Here, we write down the quantum mechanical ($0+1$ dimensional)
action for these open string degrees of freedom.
This action is a generalization of the quantum mechanical action
obtained in a solitonic approach of the baryons in holographic QCD
\cite{HSSY},
which is related to that of the collective coordinates
in the Skyrme model \cite{ANW},
and the nuclear matrix model formulated in \cite{HIY},
which is obtained by considering the ground states
in the open string spectrum.
The baryon states are obtained by quantizing this system.
In this section, we give the general procedure
to obtain the baryon spectrum including the contributions from the excited
open string states. The explicit construction of some of the low-lying
baryon states will be given in section \ref{sec:comparison}.

\subsection{The action}
The action for the open string states attached on the baryon vertex
$\dfbv$-brane is written as
\begin{eqnarray}
 S=\int dt\,(L_0+L_m)\ ,
\end{eqnarray}
where $L_0$ is the Lagrangian for the ground states
while $L_m$ is the part that involves the excited states.
$L_0$ is derived in \cite{HIY} as
\begin{eqnarray}
L_0=
\frac{M_0}{2}\left[\dot X^2
+|D_0 w|^2
-V_{\rm ADHM}(w)
-V_0(X,w)\right]+N_c A_0\ ,
\end{eqnarray}
where $w=(w_\alpha^I)$ is a complex $N_f\times 2$ matrix variable
with a spin ($SU(2)_J$) index
$\alpha=1,2$ and a flavor ($SU(N_f)_{\rm flavor}$) index $I=1,\cdots,N_f$,
$X=(X^M)$ ($M=1,2,3,z$) is a real 4 component variable,
and $A_0$ the $U(1)$ gauge field on the $\dfbv$-brane;
$w$ and $X$ correspond to the ground state for 8-4 strings
and 4-4 strings, respectively. The value of $X$ represents
the position of the $\dfbv$-brane in the 4 dimensional space
parametrized by $(x^1,x^2,x^3,z)$. The dot denotes the time
derivative as $\dot X\equiv \frac{d}{dt}X$ and
\begin{eqnarray}
D_0 w\equiv\dot w-iA_0 w\equiv \frac{dw}{dt}-iA_0 w
\end{eqnarray}
is the covariant derivative.
The potential terms are given by
\begin{eqnarray}
V_{\rm ADHM}(w)
&=&c \left(\tr(\vec\tau\, w^\dag w)\right)^2
=c\left(
2|w^\dag w|^2-(|w|^2)^2
\right)
\ ,
\label{VADHM}
\\
V_0(X,w)&=&m_z^2 (X^z)^2+\gamma |w|^2
+\frac{v}{|w|^2}\ .
\label{V0}
\end{eqnarray}
Here, $\vec\tau=(\tau^1,\tau^2,\tau^3)$ is the Pauli matrix
and we have used the notation
$|a|^2\equiv \tr(a^\dag a)=\sum_{\alpha,I}(a^\dag)^\alpha_I a_\alpha^I$
for a complex matrix $a=(a_\alpha^I)$.
$M_0$, $c$, $m_z$, $\gamma$, and $v$ are constants; $M_0$, $c$ and $m_z$
are related to the number of colors $N_c$ and the 't Hooft coupling
$\lambda$ as \footnote{
There is a mass parameter $\Mkk$ that gives the mass scale of the model.
We mainly work in the $\Mkk=1$ unit. The $\Mkk$ dependence can be easily
recovered by dimensional analysis.}
\begin{align}
M_0=\frac{\lambda N_c}{27\pi} \ ,~~
c=\frac{\lambda^2}{3^6\pi^2}\ ,~~m_z^2=\frac{2}{3}\ .
\label{constants}
\end{align}

The potential $V_{\rm ADHM}$ (\ref{VADHM}) is obtained by integrating out
the auxiliary fields in \cite{HIY}.
The condition $V_{\rm ADHM}(w)=0$ is equivalent to
the ADHM constraints for the ADHM construction
of the self-dual instanton solution.
The first term of $V_0$ in (\ref{V0}) represents the fact that the
$\dfbv$-brane is attracted to the origin in the $z$-direction due to the
curved background.
The second and third terms in  (\ref{V0}) are added rather
phenomenologically.
$\gamma$ is chosen to be $\gamma=1/6$ in \cite{HIY} so that
the second term in (\ref{V0}) recovers the corresponding term
in the soliton approach \cite{HSSY}. The third term in (\ref{V0}) was not
present in \cite{HIY}, but one could add it to have more flexibility.
We treat $\gamma$ and $v$ as unspecified parameters for the
moment.\footnote{
One motivation to add these terms is to accommodate possible additional
energy contributions from the gauge fields on the D8-branes. The second
and third terms in (\ref{V0}) mimic the $\rho$-dependent energy contributions
from the gauge fields in \cite{HSSY}. Note that we should not trust
this potential near $w=0$ when $v\ne 0$, since the third term in (\ref{V0})
diverges at $w=0$. As we will see in sections \ref{Nf2} and
\ref{LargeN}, the wavefunctions
of the baryon states that we are mostly interested in peak away from
$w=0$ and we expect that it does not affect the main features
of the analysis.
}

$L_m$ is the Lagrangian with the excited states obtained in section
\ref{sec:baryonstates}.
It can be written as
\begin{eqnarray}
L_m=\frac{M_0}{2}
\left[\sum_j
\left(|D_0\Psi_j|^2-m_j^2|\Psi_j|^2
\right)+\sum_{k}\left(\dot\Phi_k^2-m_k^2\Phi_k^2
\right)+L_{\rm int}\right]\ ,
\label{Lm0}
\end{eqnarray}
where $\Psi_j$ and $\Phi_k$
denote the fields corresponding to the excited states created
by 8-4 strings and 4-4 strings, respectively. We call these
``massive fields'' in the following.
The indices $j$ and $k$ label all the excited states and $m_j^2$ and $m_k^2$
are the mass squared of these states given in (\ref{95mass})
and (\ref{55mass}), which are of order $1/\alpha'\sim\cO(\lambda)$.
The $\Psi_j$ are complex fields that
couple with the $U(1)$ gauge field $A_0$ with the unit charge,
while $\Phi_k$ are real fields, which are neutral under the $U(1)$
gauge symmetry.
$L_{\rm int}$ gives the interaction terms for the massive fields
that may also contain massless fields.
We put the overall factor $M_0/2$ by convention so that all the fields
have the dimension of length.
Since the evaluation of the interaction terms including
the massive states is beyond the scope of this paper,
we assume that the contribution from $L_{\rm int}$ is small
as far as the qualitative features of the baryon spectrum are
concerned.
In section \ref{subsec:Lint}, we argue that though most of the possible
terms in $L_{\rm int}$ are suppressed in the large $N_c$ limit,
there are some terms that could survive even in the large
$N_c$ limit.

\subsection{Gauss law constraint and Hamiltonian}

To quantize our system, we follow the approach developed recently
in \cite{HMM}.
We take the $A_0=0$ gauge and impose the EOM for $A_0$
(Gauss law constraint) as a physical state condition on the Hilbert space.
The Gauss law constraint can be written as
\begin{eqnarray}
q_w+\sum_j q_j=N_c\ ,
\label{Glaw}
\end{eqnarray}
where
\begin{eqnarray}
q_w\equiv \frac{M_0}{2}\tr(i(\dot w^\dag w-w^\dag\dot w))
\ ,~~~
q_j\equiv \frac{M_0}{2}
i(\dot \Psi_j^\dag \Psi_j-\Psi_j^\dag\dot\Psi_j)
\ .
\end{eqnarray}
These $q_w$ and $q_j$ correspond to the charge associated with
the phase rotation symmetries $w\rightarrow e^{i\alpha_w}w$
and $\Psi_j\rightarrow e^{i\alpha_j}\Psi_j$, respectively,
which are approximate symmetries that exist when
the interaction term $L_{\rm int}$ is neglected.
The Gauss law constraint (\ref{Glaw}) represents
the fact that $N_c$ open strings have to be attached
on the $\dfbv$-brane and $q_j$ is interpreted as
the number of excited open strings associated with
$\Psi_j$.\footnote{
Both $q_w$ and $q_j$ can be negative. The sign reflects the orientation
of the fundamental string attached on the $\dfbv$-brane.
}

It is interesting to note that the Gauss law constraint
(\ref{Glaw}) implies that the spin of the baryon state is
half-integer or integer for odd or even $N_c$, respectively.\footnote{
See \cite{Hashimoto:2010rb,HMM} for related discussions.}
Indeed, the wavefunction for the baryon
state satisfying the Gauss law constraint (\ref{Glaw})
is of the form\footnote{Here, we discuss the cases with
$0\le q_w\le N_c$ for simplicity. Other cases can also be discussed
in a similar way.}
\begin{eqnarray}
 \psi(X,w,w^\dag,\Psi_j,\Psi_{j}^\dag,\Phi_k)
=\underbrace{
w^{I_1}_{\alpha_1}\cdots w^{I_{q_w}}_{\alpha_{q_w}}\Psi_{j_1}\cdots
\Psi_{j_{N_c-q_w}}
}_{N_c}
\wt\psi(X, w^\dag w,\Psi_j^\dag\Psi_{j'},\Psi_j^\dag w,w^\dag\Psi_j,\Phi_k)
\ .
\label{wf}
\end{eqnarray}
Here, $\wt\psi$ is a $U(1)$-invariant wavefunction that is written
only through $U(1)$ invariants.
Because 8-4 strings ($w$ and $\Psi_j$) and 4-4 strings
($X$ and $\Phi_k$) carry half-integer and integer spin, respectively,
$\wt\psi$ can only have an integer spin and
the spin of the state (\ref{wf}) is $N_c/2 \mod~\bZ$.

Omitting $L_{\rm int}$, the Hamiltonian in the $A_0=0$ gauge is given by
\begin{eqnarray}
 H=H_0+H_m\ ,
\end{eqnarray}
with
\begin{eqnarray}
 H_0&=&\frac{1}{2M_0}(P_X^2+|P_w|^2)+\frac{M_0}{2}(V_{\rm ADHM}(w)+V_0(X,w))\ ,
\label{H0}
\\
 H_m&=&\sum_j\left(\frac{1}{2M_0}|P_{\Psi_j}|^2+\half M_0 m_j^2|\Psi_j|^2\right)
+\sum_k\left(\frac{1}{2M_0}P_{\Phi_k}^2+\half M_0 m_k^2\Phi_k^2\right)\ ,
\label{Hm}
\end{eqnarray}
where $P_X$, $P_w$, $P_{\Psi_j}$, and $P_{\Phi_k}$ are
the momenta conjugate to $X$, $w$, $\Psi_j$ and $\Phi_k$, respectively.
$H_m$ (\ref{Hm}) is simply a collection of harmonic oscillators associated with
the excited open string states obtained in section \ref{sec:baryonstates}.
The quantum mechanics for $H_0$ (\ref{H0}) has been studied in
\cite{HIY,HMM}, though the part with $w$ is treated in a different way
in the following.

\subsection{$H_0$ for $N_f=2$}
\label{Nf2}

We are particularly interested in the cases with $N_f=2$,
in which $w$ is a $2\times 2$ complex matrix and can be
parametrized as
\begin{eqnarray}
 w=Y_0 1_2+i\vec Y\cdot\vec\tau\ ,~~~(Y=(Y_0,\vec Y)\in\bC^4)\ ,
\end{eqnarray}
where $1_2$ is the $2\times 2$ unit matrix.
$Y$ transforms as the (complex) 4 dimensional vector representation
of $SO(4)\simeq (SU(2)_I\times SU(2)_J)/\bZ_2$, where $SU(2)_J$
and $SU(2)_I=SU(N_f)_{\rm flavor}$ with $N_f=2$ corresponds
to the spin and isospin groups, respectively.
The kinetic term for $w$ in (\ref{H0}) is written as
\begin{eqnarray}
\frac{1}{2M_0}|P_w|^2= -\frac{1}{4M_0}\Delta_Y\ ,
\end{eqnarray}
where $\Delta_Y=4\frac{\del^2}{\del\ol Y_A\del Y_A}$ is the Laplacian
in $\bC^4$.

Using the relations
\begin{eqnarray}
 |w|^2= 2|Y|^2\ ,~~~
 |w^\dag w|^2=4(|Y|^2)^2-2|Y^2|^2\ ,
\end{eqnarray}
where $Y^2\equiv Y_0^2+\vec Y\cdot\vec Y$
and $|Y|^2\equiv |Y_0|^2+\vec Y^\dag\cdot\vec Y$,
the ADHM potential can be written as
\begin{eqnarray}
 V_{\rm ADHM}(Y)
=4c\left((|Y|^2)^2-|Y^2|^2\right)\ .
\end{eqnarray}
The minimum of this potential is parametrized by
\begin{eqnarray}
 Y=e^{i\theta}y\ ,~~~(\theta\in\bR\ ,~y\in\bR^4)\ .
\label{Ymin}
\end{eqnarray}
Note that $y$ together with $X$ correspond to the collective coordinates
of the one-instanton configuration considered in \cite{HSSY}.
More explicitly, 
\begin{eqnarray}
\rho\equiv \sqrt{y^2}\ ,~~~a\equiv y/\rho 
\end{eqnarray}
corresponds to the size and the $SU(2)$ orientation of the
instanton solution, respectively.\footnote{
Using the relation $a^2=1$, one can show that
$\mbox{\boldmath $a$}\equiv a_01_2+i\vec a\cdot\vec\tau$ is an element of $SU(2)$.
This $\mbox{\boldmath $a$}$ is also related to the collective coordinate of
the Skyrmion for $N_f=2$.\cite{ANW}
}
One way to include the components that are orthogonal to the
directions along (\ref{Ymin}) is to parametrize $Y$ as\footnote{
The notation $y$ and $\wt y$ in this section should not be confused with
that in section \ref{subsec:states}.}
\begin{eqnarray}
 Y=e^{i\theta}(y+i\wt y)\ ,~~~(\theta\in\bR\ ,~y,\wt y\in\bR^4)
\label{Y}
\end{eqnarray}
with
\begin{eqnarray}
\wt y=\beta_a i\Sigma^a a\ ,~~~
((\beta_a)= (\beta_1,\beta_2,\beta_3)\in\bR^3)\ ,
\end{eqnarray}
where $\Sigma^a$ ($a=1,2,3$) are the generators of $SU(2)_I$ acting on
$y$, which are chosen to be pure imaginary anti-symmetric matrices.
See Appendix \ref{SO4} for the explicit forms. One can easily
show that
\begin{eqnarray}
 y\cdot\wt y=0\ ,~~~\wt y^2=\beta^2\ ,
\end{eqnarray}
where $\beta^2=\beta_a\beta_a$ and the ADHM potential becomes
\begin{eqnarray}
 V_{\rm ADHM}(Y)=16c\, \rho^2\beta^2\ .
\label{VADHMrhobeta}
\end{eqnarray}

Note that the parametrization (\ref{Y}) has a redundancy
induced by the $\bZ_2$ transformation
\begin{eqnarray}
 \theta\rightarrow \theta+\pi\ ,~~y\rightarrow -y\ .
\label{Z2tr}
\end{eqnarray}
When the wavefunction is written in terms of $\theta$,
$y$ and $\beta_a$ instead of $Y$, we should impose the
invariance of the wavefunction under this $\bZ_2$ transformation.

In this paper, we consider the cases that $\beta$ takes
small values so that $V_{\rm ADHM}$ does not generate an additional mass
term for $\rho$.
One important observation is that the kinetic term of the Hamiltonian
(\ref{H0}) contains a term as
\begin{eqnarray}
-\frac{1}{2M_0\rho^2}\frac{\del^2}{\del\theta^2}\ ,
\label{ddtheta}
\end{eqnarray}
for $\beta^2\ll\rho^2$. (See (\ref{Laplacian}))
Since $q_w$ is the generator
of the phase rotation of $Y$, we have the relation
\begin{eqnarray}
 q_w=-i\deldel{}{\theta}
\label{qw}
\end{eqnarray}
in the quantum mechanics.
When we consider the cases with $\sum_j q_j\sim\cO(1)$,
$q_w$ has to be of $\cO(N_c)$ because of the Gauss law constraint
(\ref{Glaw}). In such cases, the term (\ref{ddtheta}) gives
a potential of the form
\begin{eqnarray}
-\frac{1}{2M_0\rho^2}\frac{\del^2}{\del\theta^2}
\sim \frac{N_c}{\lambda\rho^2}\ ,
\label{Nclamrho}
\end{eqnarray}
up to a numerical factor in the large $N_c$ limit, which has
the effect of pushing $\rho$ to have a larger value.
Let $\rho_0$ be the value of $\rho$ that minimizes
the effective potential given by adding this term to $V_0$
(\ref{V0}). Assuming that the third term in (\ref{V0}) is
either negligible or of the same order as (\ref{Nclamrho}), i.e.
$v\sim \cO(\lambda^{-2})$, we find $\rho_0^2\sim\cO(\lambda^{-1})$,
which is consistent with the results in \cite{Hong:2007kx,HSSY}.
We will shortly obtain an explicit expression
for $\rho_0$ in the large $N_c$ limit (see (\ref{rho0})),
and show that it has the effect of generating a large
mass term for $\beta_a$ in the next subsection.

\subsection{Large $N_c$ limit}
\label{LargeN}

Now, let us figure out which terms in $H_0$ are important in the large
$N_c$ limit.
First we decompose $\rho$ as $\rho=\rho_0+\drho$, and
regard $M_0^{1/2}\drho$, $M_0^{1/2}\beta_a$, and $a$ to be
order 1 variables,\footnote{This is equivalent to writing down
the Lagrangian in terms of the canonically normalized fields
$\wt\drho\equiv M_0^{1/2}\drho$ and $\wt\beta_a\equiv M_0^{1/2}\beta_a$
and taking the large $N_c$ limit with these fields kept finite.
On the other hand, $a$ satisfies $a^2=1$ by definition and
hence we regard it as an order 1 variable. We also assume here
that quantum numbers for the baryon state such as spin and isospin
are all order 1, except for $q_w$ which is assumed to be of order $N_c$
as discussed around (\ref{qw}).
}which means that
\begin{eqnarray}
 \drho\sim\beta_a\sim\cO(\lambda^{-1/2}N_c^{-1/2})\ ,~~~
 \deldel{}{\rho}\sim\deldel{}{\beta_a}\sim 
 \cO(\lambda^{1/2}N_c^{1/2})\ .
\end{eqnarray}
Then, the leading ($\cO(\lambda N_c^2)$) and subleading
($\cO(\lambda N_c)$) terms in the Laplacian $\Delta_Y$ turn out to be
\begin{eqnarray}
\Delta_Y\simeq
\left(\frac{1}{\rho^2}
+\frac{3\beta^2}{\rho_0^4}
\right)\deldel{{}^2}{\theta^2}
+\deldel{{}^2}{\rho^2}+\left(\deldel{}{\beta_a}\right)^2
+\cO(\lambda N_c^{1/2})\ .
\label{Laplacian}
\end{eqnarray}
Keeping these terms, the Hamiltonian for $\rho$ and $\beta_a$
becomes
\begin{eqnarray}
H_0|_{\rho,\beta} &\simeq&
\frac{1}{4M_0}\left[
\frac{q_w^2}{\rho^2}+\frac{3q_w^2}{\rho_0^4}\beta^2
-\deldel{{}^2}{\rho^2}-\left(\deldel{}{\beta_a}\right)^2
\right]
+\frac{M_0}{2}\left[16c\,\rho^2\beta^2+2\gamma(\rho^2+\beta^2)
+\frac{v}{2(\rho^2+\beta^2)}
\right]
\nn\\
&\simeq&
2M_0\gamma\rho_0^2
-\frac{1}{4M_0}\left[
\deldel{{}^2}{\rho^2}+\left(\deldel{}{\beta_a}\right)^2
\right]
+M_0\left(\omega_{\drho}^2\drho^2
+\omega_\beta^2\beta^2\right)\ ,
\label{H0rhobeta}
\end{eqnarray}
where
\begin{eqnarray}
\omega_{\drho}^2=4\gamma\ ,~~~
\omega_\beta^2=8c\rho_0^2+\gamma-\frac{v}{4\rho_0^4}+\frac{3q_w^2}{4M_0^2\rho_0^4}\ .
\label{omega}
\end{eqnarray}
Here we have imposed the
condition that $\rho_0$ minimizes the potential
for $\rho$, which reads
\begin{eqnarray}
\rho_0^2=\half\sqrt{\frac{1}{\gamma}\left(\frac{q_w^2}{M_0^2}+v\right)}
\ .
\label{rho0}
\end{eqnarray}
The Hamiltonian (\ref{H0rhobeta}) is a sum of
the harmonic oscillators for $\rho$ and $\beta_a$.

A few comments are in order:
First, $\omega_{\drho}^2$ coincides with $m_z^2$ in (\ref{constants})
for $\gamma=1/6$ used in \cite{HIY}, which is consistent with
\cite{HSSY}. Second, the value of $\rho_0$ in (\ref{rho0}) agrees with
that in \cite{HIY} when $q_w=N_c$ and $v=0$. However, as pointed
out in \cite{HIY}, it is larger than the value in
\cite{Hong:2007kx,HSSY} by a factor of $5/4$. One can adjust the value
of $v$ as $v=-\frac{N_c^2}{5M_0^2}$ to match with the value in
\cite{Hong:2007kx,HSSY}.
Third, on the right-hand side of $\omega_\beta^2$ in (\ref{omega}),
the first term $8c\rho_0^2$ is of order $\lambda$,
while the other terms are of order 1. Recall that the masses of the
excited open string states 
are $m^2\propto 1/\alpha'\sim\cO(\lambda)$. This means that although
$\beta_a$ arises as the ground states
(the open string states with $N_{95}=0$),
it acquires a large mass comparable to the massive excited states due to
the ADHM potential (\ref{VADHM}) together with the Gauss law constraint
(\ref{Glaw}).

\subsection{Mass formula}
\label{bmassf}

As argued in section \ref{Nf2},
the Hamiltonian is reduced to a collection of harmonic oscillators
in the large $N_c$ limit, which can be easily solved.
Then, the masses of the baryons are obtained as
\begin{eqnarray}
 M= M_0^*
+m_z n_z+\omega_{\drho}n_\rho+
\omega_\beta\sum_{a=1}^3 n_\beta^a
+\sum_{j}m_j (n^{\Psi}_j+n^{\ol\Psi}_j)+\sum_{k}m_k n^{\Phi}_k\ ,
\label{mass}
\end{eqnarray}
where $n_z$, $n_\rho$, $n_\beta^a$, $n^{\Psi}_j$, $n^{\ol\Psi}_j$, and
$n^{\Phi}_k$ are non-negative integers corresponding to the
excitation levels of the harmonic oscillators associated with
$X^z$, $\drho$, $\beta_a$, $\Psi_j$, $\ol\Psi_j$, and $\Phi_k$,
respectively; $m_z$, $\omega_{\drho}$, and $\omega_{\beta}$ are given in
(\ref{constants}) and (\ref{omega}); $m_j$ and $m_k$ are
the masses for the corresponding open string states given in
(\ref{95mass}) and (\ref{55mass}), respectively, and
$M_0^*$ is a ($q_w$-dependent) constant whose classical value is
\begin{eqnarray}
 M^*_{0\,{\rm classical}}=(1+2\gamma\rho_0^2)M_0\ ,
\label{M0cl}
\end{eqnarray}
where the first term $M_0$ comes from the tension of the $\dfbv$-brane
placed at $y=z=0$ and the second term $2\gamma\rho_0^2M_0$ is the first
term in (\ref{H0rhobeta}).
It also contains the contributions from the zero-point energies of all
the fields in the system, including those neglected in section
\ref{sec:baryonstates}. Since there are infinitely many fields involved,
it is not easy to evaluate it explicitly.\footnote{
As pointed out in \cite{HSSY}, a similar problem also appears in the
soliton approach.} For this reason, we leave $M_0^*$ as an unknown parameter
and focus on the mass differences.

Note that the mass (\ref{mass}) implicitly depends on the value of $q_w$
through the parameters $M_0^*$ and $\omega_\beta$. Because the  
Gauss law constraint (\ref{Glaw}) implies that $q_w$ is related to
$n_j^\Psi$ and $n_j^{\ol\Psi}$ by
\begin{eqnarray}
 q_w+\sum_j(n^{\Psi}_j-n^{\ol\Psi}_j)=N_c\ ,
\end{eqnarray}
these parameters are state dependent.

As a consistency check, one can show
that the formula (\ref{mass}) agrees with the leading-order
terms in the baryon mass formula obtained in \cite{HSSY}
when $q_w=N_c$ and $n_\beta=n_j^{\Psi}=n_j^{\ol\Psi}=n_k^\Phi=0$.
In fact, the baryon mass formula in \cite{HSSY} can be written as
\begin{eqnarray}
M&=&
M_0+\sqrt{\frac{(\ell+1)^2}{6}+\frac{2}{15}N_c^2}
+\frac{2(n_\rho+n_z)+2}{\sqrt{6}}
\label{HSSYmass}
\\
&\simeq&
M_0+2M_0\gamma\rho_0^2
+\frac{(\ell+1)^2}{4M_0\rho_0^2}
+\omega_{\drho} n_\rho +m_z n_z+\half(\omega_{\drho}+m_z)
+\cO(N_c^{-3})\ ,
\label{HSSYmass2}
\end{eqnarray}
where $\ell\in\bZ_{\ge 0}$ is related to the spin $J$ and isospin $I$
as $I=J=\ell/2$.
The $\ell$ dependence appears because the Laplacian in the $y$-space:
\begin{eqnarray}
 \Delta_y\equiv\left(\deldel{}{y_A}\right)^2
=\frac{1}{\rho^3}\del_\rho(\rho^2\del_\rho)+\frac{1}{\rho^2}
\Delta_{S^3}\ ,
\end{eqnarray}
contains the Laplacian on $S^3$ parametrized by $a$, denoted
by $\Delta_{S^3}$, whose eigenvalue is $-\ell(\ell+2)$.
In (\ref{Laplacian}), we have neglected this contribution,
though it also appears in $\Delta_Y$ if we keep the
$\cO(N_c^0)$ term.

In \cite{HSSY}, $\ell$ was chosen to be odd (or even)
for odd (or even) $N_c$ by hand, so that the spin of the baryon
obtained in the soliton approach is consistent
with that in the quark model, as it is also the case
for the Skyrme model with $N_f=2$.
In our case, this condition is replaced with
$\ell\equiv q_w~({\rm mod}~2 )$, which automatically
follows from the fact that the eigenfunction
of $\Delta_{S^3}$ is given by
\begin{eqnarray}
T^{(\ell)}(a)\equiv 
C^{A_1\cdots A_\ell}a_{A_1}\cdots a_{A_\ell}\ ,
\label{Tell}
\end{eqnarray}
where  $C^{A_1\cdots A_\ell}$ is a traceless
symmetric tensor of rank $\ell$, and $\theta$
appears in the wavefunction as an overall factor $e^{iq_w\theta}$.
As explained around (\ref{Z2tr}), the wavefunction has to be
invariant under the $\bZ_2$ transformation (\ref{Z2tr}), which implies
$\ell\equiv q_w~({\rm mod}~2 )$.

\subsection{Wavefunctions of the baryon states}
\label{wavefunc}

As discussed above, the Hamiltonian of the one-baryon quantum
mechanics is a collection of infinitely many harmonic
oscillators in the large $N_c$ limit.
The eigenfunction can be written as a product of
a function of $X$, $a$, $\drho$ and
$\beta_a$, and a function of $\Psi_j$, $\Psi_j^\dag$ and
$\Phi_k$ as
\begin{eqnarray}
 \psi(X, a, \drho,\beta_a,\Psi_j,\Psi_j^\dag,\Phi_k )=
\psi_0(X, a, \drho,\beta_a)\,\psi_m(\Psi_j,\Psi_j^\dag,\Phi_k )\ .
\end{eqnarray}
We call $\psi_0$ and $\psi_m$ wavefunctions
for the massless and massive sectors, respectively.\footnote{
Although $X^z$, $\drho$ and $\beta_a$ have mass terms in the Hamiltonian
(\ref{H0}) and (\ref{H0rhobeta}), we consider them to be
in the massless sector, because these modes originate from the
the massless open string states in the flat spacetime limit.
}

The massless sector wavefunction $\psi_0$ can be written as
\begin{eqnarray}
\psi_0= e^{i\vec p\cdot\vec X} 
T^{(\ell)}(a)\psi_{n_z}(X^z)\psi_{n_\rho}(\drho)\psi_{n_\beta}(\beta_a)\ ,
\end{eqnarray}
where $e^{i\vec p\cdot \vec X}$ is the wavefunction for the
plane wave with momentum $\vec p$,
$T^{(\ell)}(a)$ is defined in (\ref{Tell}), and
$\psi_{n_z}$, $\psi_{n_\rho}$ and $\psi_{n_\beta}$
are the eigenfunctions of the harmonic oscillators
for $X^z$, $\drho$ and $\beta_a$ with
the excitation numbers $n_z$, $n_\rho$ and $n^a_\beta$, respectively.
We set $\vec p=0$ in the following for simplicity.
We also use the bra-ket notation as
\begin{eqnarray}
\ket{\psi_0}=\ket{\ell,n_z,n_\rho,n^a_\beta,q_w}\ .
\end{eqnarray}
Here, $q_w$ is included in the notation to remember that the massless
sector wavefunction also depends on $q_w$.

If $\psi_{n_\beta}$ is trivial, $\psi_0$ agrees with the large $N_c$
limit of the wavefunction obtained in \cite{HSSY}.
As shown in \cite{HSSY}, $T^{(\ell)}(a)$ has a degeneracy
of $(\ell+1)^2$ that corresponds to the states in the representation
of $I=J=\ell/2$.
The mass formula (\ref{mass}) appears to be independent of $\ell$,
because the $\ell$ dependence is a subleading effect in the large $N_c$
limit. Upon taking finite $N_c$ effects into account, 
we expect that the energy is an increasing function of $\ell$ as
in \cite{HSSY}.\footnote{
If we set $N_c=3$ in the mass formula (\ref{HSSYmass}) given in
\cite{HSSY}, the expansion as (\ref{HSSYmass2}) is not justified
for $\ell>1$. This suggests that the $\ell$ dependence is actually
important to compare with the realistic QCD. (See \cite{HSSY} for
further discussion.)}

Note that since $X^z$ is parity odd, $\psi_{n_z}$ has parity
$(-1)^{n_z}$. As mentioned in section \ref{LargeN},
$\omega_{\drho}$ coincides with $m_z^2$ for $\gamma=1/6$
and hence the states with $(n_\rho,n_z)=(1,0)$ and
$(n_\rho,n_z)=(0,1)$ are degenerate. This implies
a degeneracy between parity even and odd states
for those with $(n_\rho,n_z)\ne (0,0)$.
This could be a hint toward an understanding of the
parity doubling phenomenon in the excited baryons.\footnote{
See, e.g., \cite{Afonin:2007mj} for a review.}

$\psi_{n_\beta}$ is a wavefunction for
a 3 dimensional harmonic oscillator with respect to $\beta_a$
($a=1,2,3$). The energy contribution in the mass formula
(\ref{mass}) for this part is $\omega_\beta n_\beta$ with
\begin{eqnarray}
 n_\beta\equiv \sum_{a=1}^3 n_\beta^a\ .
\end{eqnarray}
The degeneracy is
\begin{eqnarray}
 \half(n_\beta+1)(n_\beta+2)\ ,
\end{eqnarray}
and the eigenspace for a given $n_\beta$ can be decomposed
into a direct sum over the states
with isospin $I=0,2,\cdots,n_\beta$
or $I=1,3,\cdots,n_\beta$ for even or odd $n_\beta$,
respectively.
For example, for the state with $\ell=1$ and $n_\beta=1$,
the massless wavefunction
$\psi_0$ has spin $1/2$ and isospin $1/2\otimes 1=3/2\ds 1/2$.

The wavefunction for the massive sector is given by
the eigenfunctions of the harmonic oscillators associated
with $\Psi_j$, $\Psi_j^\dag$ and $\Phi_k$, which is written
in the bra-ket notation as
\begin{eqnarray}
\ket{\psi_m}=\ket{n_j^\Psi,n_{j}^{\ol\Psi},n_k^\Phi}\ .
\end{eqnarray}
In order to classify these states, we introduce the notation
\begin{eqnarray}
 \cN=\cN_{84}+\cN_{44}\ ,
\label{cN}
\end{eqnarray}
which we call the level of a baryon, with
\begin{eqnarray}
 \cN_{84}=\sum_{j}(n_j^\Psi+n_j^{\ol\Psi}) N_{84}^{(j)}\ ,~~~
 \cN_{44}=\sum_{k} n_k^\Phi N_{44}^{(k)}\ ,
\end{eqnarray}
where $N_{84}^{(j)}$ and $N_{44}^{(k)}$ are
the excitation numbers for $\Psi_j$ and $\Phi_k$ given in
Table \ref{table:95summary} and Table \ref{table:55summary},
respectively.\footnote{$N_{84}$ and $N_{44}$ correspond
to $N_{95}$ and $N_{55}$ in section \ref{sec:baryonstates},
respectively.
}
It will become increasingly complicated to extract
the spin and isospin for the states with larger $\cN$.
We will give some explicit examples of the baryon states
in section \ref{sec:comparison}.

\subsection{Comments on $L_{\rm int}$}
\label{subsec:Lint}

Here, we make some comments on $L_{\rm int}$ in (\ref{Lm0}).
First, we classify
$L_{\rm int}$
depending on the order of the massive fields multiplied and
assume that each term contains at least two massive fields so that
the trivial configuration $\Psi_j=\Phi_k=0$ is a solution of
the EOM for the massive fields.
Note that the overall factor $M_0$ in the
Lagrangian (\ref{Lm0}) is proportional to $N_c$, which reflects the
fact that the leading terms of the open string
action are given by the string worldsheet of disk topology.
As always, we neglect the loop corrections of string theory
which are suppressed by $1/N_c$. Then, $L_{\rm int}$ is order 1
in the $1/N_c$ expansion with fixed $\lambda$.
If one writes down the Lagrangian using canonically normalized
massive fields
\begin{eqnarray}
\wt\Psi_j\equiv \sqrt{M_0}\Psi_j\ ,~~~
\wt\Phi_k\equiv \sqrt{M_0}\Phi_k\ ,
\end{eqnarray}
one finds that all the terms with more than two massive fields
are suppressed in the large $N_c$ limit. Therefore,
the terms in $L_{\rm int}$ that survive in the large $N_c$ limit are quadratic
with respect to the massive fields. For the same reason,
it should not contain $\dot w$, $\dot X$ or $X$. Then, the possible terms 
consistent with the $U(1)$ gauge symmetry are schematically written as
\begin{eqnarray}
w^n (w^\dag)^n \Psi_j^\dag \Psi_{j'}\ ,~~~
w^n (w^\dag)^n \Phi_k^\dag \Phi_{k'}\ ,~~~
w^n (w^\dag)^{n+2} \Psi_j \Psi_{j'}\ ,~~~
w^n (w^\dag)^{n+1} \Psi_j \Phi_k\ ,
\label{wPP}
\end{eqnarray}
with properly contracted indices
and their complex conjugates. As we have seen in sections \ref{Nf2}
and \ref{LargeN}, $w$ is treated as an order 1 variable and these terms
may appear even in the large $N_c$ limit.

One might think that these terms are perhaps suppressed for large $\lambda$.
Unfortunately, however, the answer is no. Consider, for example,
a term proportional to $|w|^{2n}|\Psi_j|^2\propto |Y|^{2n}|\Psi_j|^2$
for $N_f=2$. As observed in section \ref{LargeN}, the leading term in
$Y$ is $Y\sim \rho_0 a\sim \cO(\lambda^{-1/2})$. Recall that all the
fields have the dimension of length in our convention. To have the
correct dimensions, there should be an appropriate number of $\alpha'$ or
$\Mkk$ in the coefficient of (\ref{wPP}) to saturate the correct
dimension of $L_{\rm int}$. A possible term is of the form
\begin{eqnarray}
L_{\rm int}
\sim \alpha'^{-n-1}|Y|^{2n}|\Psi_j|^2
\sim \lambda |\Psi_j|^2\ ,
\label{Lint}
\end{eqnarray}
which shifts the mass for $\Psi_j$
in the same order as the original mass term. This is the same mechanism
as the mass generation of $\beta_a$ discussed in section \ref{LargeN}.
$L_{\rm int}$ may also induce mixing terms as well, and the
diagonalization of the mass matrix may become very complicated.
Because we do not know the explicit form of $L_{\rm int}$, we are not able
to evaluate it explicitly and leave the detailed analysis including
$L_{\rm int}$ for future research.

%
%%%%%%%%%%%%%%%%%%%%%%%%%%%%%%%%%%%%%%%%%%%%%%%%%%%%%%%%%%%%%%%%%%%
%
\section{Comparison with experiments}
\label{sec:comparison}

\subsection{Regge trajectory}
\label{Regge}

Here, we focus on the baryons listed in Table \ref{Nregge},
which are the lightest baryons with $I=1/2$
and $J^P=(n+1/2)^{(-)^n}$ ($n=0,1,\cdots,5$) found in the experiments.
\begin{table}[H]
\begin{center}
\begin{tabular}{c||c|c|c|c|c|c}
 baryons & N & N(1520)& N(1680)& N(2190)& N(2220)& N(2600)%& N(2700)
\\\hline
$J^P$ & $1/2^+$ & $3/2^-$& $5/2^+$& $7/2^-$& $9/2^+$& $11/2^-$%& $13/2^+$
\\\hline
mass[MeV]& 939& 1510$\sim$1520 & 1680$\sim$1690&2140$\sim$2220&
 2250$\sim$2320 & 2550$\sim$2750%& 2700
\end{tabular}
\parbox{15cm}{
\caption{Nucleon and lightest baryons with $I=1/2$
and $J^P=(n+1/2)^{(-)^n}$ ($n=0,1,\cdots,5$).
Data taken from the baryon summary table in \cite{pdg}}
\label{Nregge}}
\end{center}
\end{table}

These baryons have been considered to be described by an excited
(rotating) open string with a quark and diquark pair attached on the
two end points.\cite{Sharov:2013tga,cobi}\footnote{For earlier and
closely related works, see \cite{3strings}. See also \cite{diquark} for
related works based on quark-diquark models.
}
An analogous object in our model is a $\dfbv$-brane
with $(N_c-1)$ 8-4 strings in the ground state and only one 8-4 string
being excited as $J$ increases. The aim of this subsection
is to discuss whether our model gives us plausible predictions
assuming that this is the correct interpretation.
More explicitly, the lightest one in Table \ref{Nregge},
which is the nucleon (proton or neutron), is identified
with $q_w=N_c$, $\ell=1$ and
$n_\rho=n_z=n_j^\Psi=n_j^{\ol\Psi}=n_k^\Phi=0$.\footnote{
Here, we consider $N_c$ to be a large odd number.
Recall that the condition $\ell\equiv q_w$ (mod 2) has to be satisfied.
(See section \ref{bmassf}.)
}
The excited nucleons with spin $J\ge 3/2$ in Table \ref{Nregge}
are interpreted as the highest spin state among those
with $q_w=N_c-1$, $\ell=0$,
$n_\rho=n_z=n_{j'}^{\ol\Psi}=n_k^\Phi=0$ and $n_{j'}^\Psi=\delta_{j'j}$
for some $j$.
These states are most likely to be the lightest
state among the highest spin states with isospin 1/2 for each level. 
Let us discuss if the quantum numbers and the masses of these states are
consistent with the experimental data with this interpretation.

The states we consider are labeled uniquely by the level $\cN$
introduced in (\ref{cN}).
Let $M_\cN$ denote the baryon mass for a given $\cN$.
The nucleon corresponds to the case $\cN=0$, which has
$J^P=1/2^+$ and mass given by
\begin{align}
 M_{\cN=0}=M_0^\ast(q_w=N_c,\ell=1) \ .
\label{EN0}
\end{align}
Here, $M_0^\ast$ is considered to be a function of $q_w$ and $\ell$
as argued in section \ref{bmassf}.
As it is technically hard to compute the quantum $M_0^\ast(q_w,\ell)$,
we regard it as an unknown parameter.

{}For $\cN\ge 1/2$, because $\ell=0$, the massless sector
has vanishing spin and isospin.
Then, the total spin of the excited baryons with $\cN\ge 1/2$
is fixed by the massive sector.
Let the excitation number of the excited 8-4 string be $N^{(j)}_{84}$,
which is to be identified with the level $\cN$ for
the excited nucleons as seen before.
{}For each $\cN=1/2,1,3/2,2,\cdots$, the highest spin states 
are contained in the states of the form
\begin{align}
  \left(
\alpha_{-1/2}^{M_1}\cdots \alpha_{-1/2}^{M_{2\cN}}
-(\mbox{trace parts})\right)\ket{a,I}_{\NS} \ ,
\label{highest59}
\end{align}
which belongs to the spin $(\cN,\cN)\otimes (1/2,0)$ representation
of $SU(2)_L\times SU(2)_R$.
Here, we have included the flavor
index $I$ to show that it is an isospin $1/2$ state for $N_f=2$.
Decomposing this under the vector-like subgroup
$SU(2)_{J}\subset SU(2)_L\times SU(2)_R$, 
one finds that the highest spin is given by $J=2\cN+1/2$.
The parities of these excited nucleons are given by
$P=(-)^{2\cN}$, 
because the state (\ref{highest59}) has parity $(-)^{2\cN}$
and the massless sector is parity even for $n_z=0$.
Therefore, the spin, isospin and parity for the excited nucleon states
constructed above are consistent with those in Table \ref{Nregge}.

The baryon mass formula (\ref{mass}) implies that the masses for
these excited nucleons with $J\ge 3/2$ states are
\begin{align}
M_{\cN}=
M_0^\prime+\sqrt{\frac{\cN}{\alpha'}}
=
M_0^\prime+\frac{1}{\sqrt{2\alpha^\prime}}
\sqrt{J-\frac{1}{2}} \ ,
\label{EN}
\end{align}
where, $M_0^\prime\equiv M_0^\ast(q_w=N_c-1,\ell=0)$.
This formula can be recast as a formula for spin $J$
as a function of mass $M$:
\begin{align}
J=2\alpha^\prime(M-M_0^\prime)^2+\frac{1}{2} \ .
\label{JE}
\end{align}

\begin{figure}[htbp]
\begin{center}
%\leavevmode
%\psbox[height=8cm]{fig1.eps}
\includegraphics[scale=1]{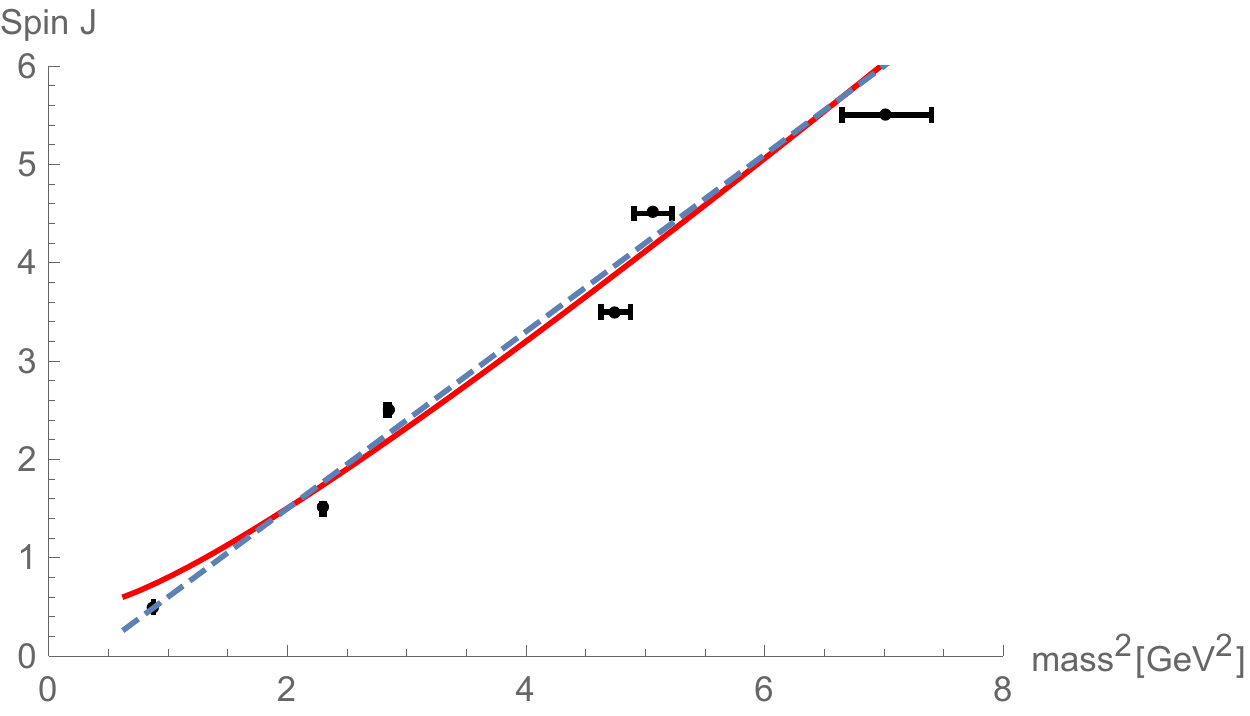}
\parbox{15cm}{
\caption{
A plot of (\ref{JE}) compared with the experimental data.
The dots with error bars represent the data listed in Table
 \ref{Nregge}.
The solid line is the plot of (\ref{JE}) with
$\alpha'\simeq 0.6\,{\rm GeV}^{-2}$ and $M_0'\simeq 0.5\,{\rm GeV}$,
while the dashed line is the linear trajectory (\ref{linear}) with
$\alpha_0 \simeq -0.3$ and $\alpha^\prime\simeq 0.9\,{\rm GeV}^{-2}$.
}\label{plot}}
\end{center}
\end{figure}

It has been observed that, when the spin $J$ is plotted
as a function of the mass squared $M^2$, the excited nucleon states
listed in Table \ref{Nregge} lie on a linear trajectory that satisfies
\begin{align}
  J=\alpha_0+\alpha^\prime M^2 \ .
\label{linear}
\end{align}
with $\alpha_0|_{\rm exp}\simeq -0.3$ and
$\alpha^\prime|_{\rm exp}\simeq 0.9\,{\rm GeV}^{-2}$.
Our formula (\ref{JE}) is a nonlinear function with respect to $M^2$,
and one would think it disagrees with the observation. However, choosing
\begin{eqnarray}
\alpha'\simeq 0.6\,{\rm GeV}^{-2}\ ,~~~M_0'\simeq 0.5\,{\rm GeV}\ ,
\label{alpha_m0}
\end{eqnarray}
we get the plot shown in Figure \ref{plot}, which shows that it can fit
the data reasonably well.
%%%%%%%%%%%%
%%%%%%%%%%%%
%
Due to the nonlinear term in (\ref{JE}), the trajectory in Figure
\ref{plot} is curved toward the left and the value of mass squared
for $J=1/2$ becomes significantly smaller compared to that of the
nucleons (proton or neutron). This is, however, not a problem
of the formula (\ref{JE}) as it is derived for the states with
$J\ge 3/2$. Our expression for the nucleon mass is given in (\ref{EN0}).
Though we are not able to predict its value, this observation suggests
that the difference between (\ref{EN0}) and $M_0'$
\begin{eqnarray}
 M_{\cN=0}-M_0'=
 M_0^*(q_w=N_c,\ell=1)-M_0^*(q_w=N_c-1,\ell=0) \ ,
\end{eqnarray}
is positive, as expected.\footnote{
To get a rough estimate, one could try to evaluate it by
assuming that $\ell$ dependence is small and the mass difference
$\Delta M_0^*\equiv  M_{\cN=0}-M_0'$ is entirely determined by
(\ref{M0cl}). Then, one gets
$\Delta M_0^*=2 M_0\gamma(\rho_0^2|_{q_w=N_c}-\rho_0^2|_{q_w=N_c-1})$.
For $\gamma=1/6$ and $v=0$, using (\ref{rho0}), we get
$\Delta M_0^*=\Mkk/\sqrt{6}$, where we have recovered the $\Mkk$
dependence by dimensional analysis.
Using the value of $\Mkk$ in (\ref{mkklambda}), this is estimated as
387 MeV.
}

We emphasize that the values (\ref{alpha_m0}) should not be considered
to be an accurate estimate, because we have neglected all the $1/N_c$
and $1/\lambda$ corrections, as well as the possible contributions from
the interaction term (\ref{Lint}) for the massive fields.
Nevertheless, let us make a few comments here on the value of
$\alpha'$. In \cite{SS1,SS2}, the parameters $\Mkk$ and $\lambda$ were
chosen to be
\begin{align}
  \Mkk\simeq 949\,{\rm MeV}\ ,~~\lambda\simeq 16.6 \ ,
\label{mkklambda}
\end{align}
to fit the experimental values of the $\rho$-meson mass and
the pion decay constant. If we use these values and the relation
(\ref{lslam}), we obtain
$\alpha^\prime\simeq 0.452\,{\rm GeV}^{-2}$,
which is a bit small compared with the value in (\ref{alpha_m0}).
On the other hand, the value of $\alpha'$  evaluated from the
Regge slope of the $\rho$-meson trajectory is
$\alpha'|_{\rm exp}\simeq 0.88~{\rm GeV}^{-2}$.
In \cite{HSS}, the $\rho$-meson Regge behavior is analyzed theoretically
using the same holographic model of QCD as in the present paper. It was
argued there that the $\rho$-meson trajectory has some nonlinear
corrections similar to that in (\ref{JE}) and the value of $\alpha'$
that fits well with the experimental data turned out to be around
$1.1~{\rm GeV}^{-2}$.
The value of $\alpha'$ in (\ref{alpha_m0}) is close to neither of these
values, though it is not too far from them.
It is important to resolve this discrepancy by
making more accurate estimate of $\alpha^\prime$.

Note that the slope $\alpha'$ of the linear Regge trajectory
(\ref{linear}) for the excited nucleons is very close to that
of the $\rho$-mesons.
This is one of the motivations for conjecturing that both of them
are described by open strings with some particles attached
on the end points as investigated in \cite{Sharov:2013tga,cobi}.
Our description is similar to these models in that
only one of $N_c$ strings attached on the baryon vertex gets excited
while the rest remain in the ground state.
This system may be approximated with a single open string by 
regarding the effect of the baryon vertex as a massive end point.
However, a clear distinction from the models in \cite{Sharov:2013tga,cobi}
is that the mass of the end point in the present model
is of $\cO(N_c)$ and considered to be much heavier than
the energy scale determined by the string tension.
In fact, it is not difficult to verify that a rotating
open string with a massive end point of mass $M_0$ has a classical
energy $E$ that reduces in the heavy end point limit to
\begin{align}
  J=2\alpha^\prime(E-M_0)^2 \ ,
\label{JEcl}
\end{align}
which agrees with (\ref{JE}) up to an additive constant $1/2$
and the contributions from the zero-point energy in $M_0'$.\footnote{
{}For a systematic treatment of the classical motion of rotating strings
with massive end points, see the third paper in \cite{cobi}.}
We note that the difference between the mass formula (\ref{JE})
and (\ref{JEcl}) is due to quantum $1/N_c$ corrections.

\subsection{More about excited  baryon states}
%
%\label{sec:baryon}
%

In this subsection, we show some examples of low-lying excited
baryons that are obtained in a manner explained in the
previous sections. 
{}For simplicity, we set $(n_\rho,n_z)=(0,0)$ and $n_\beta^a=0$.
The states in the massless and massive sectors are denoted by
$|\ell,q_w\rangle$ and 
$|n_j^\Psi,n_j^{\ol\Psi},n_k^\Phi\rangle$, respectively,
where only nonvanishing quantum numbers are indicated explicitly
for notational simplicity.

We start from the sector $\cN=1/2$.
This sector is constructed only from the excitation
of a single 8-4 string with $N_{84}^{(1)}=1/2$,
because any 4-4 excited state has $N_{44}^{(k)}\ge 1$.\footnote{
$N^{(j)}_{84}$ and $N^{(k)}_{44}$ are the excitation numbers for
$\Psi_j$ and $\Phi_k$, respectively. $\Psi_j$ with $j=1,2,3,4$
and $\Phi_k$ with $k=1,2,3,4$ are listed in Tables \ref{table:95summary}
and \ref{table:55summary}, respectively.
}
The corresponding field $\Psi_1$ belongs to
$(1,1/2)_-$ under $SU(2)_L\times SU(2)_R$ with the subscript denoting parity,
and yields a harmonic oscillator with angular frequency
given by $m_1=\sqrt{1/(2\alpha')}$. The condition $\cN_{84}=1/2$
is satisfied when $(n_1^\Psi,n_1^{\ol \Psi})=(1,0)$ or $(0,1)$.
The excited states $|n_1^\Psi=1\rangle$
and $|n_1^{\ol \Psi}=1\rangle$
have the same energy eigenvalue of $H_m$ with $SU(2)_L\times SU(2)_R$ 
spin given by $(1,1/2)_-$.
We consider only the former, because
this leads to $q_w=N_c-1$ so that 
(\ref{M0cl}) shows that the corresponding massless sector
has less energy compared to that with $q_w=N_c+1$,
which corresponds to $|n_1^{\ol \Psi}=1\rangle$.
As $q_w$ is even, the massless sector is allowed to have
$\ell=0,2,4,\cdots$.
We first consider the case $\ell=0$, which yields the lightest state
in the massless sector $\ket{\ell=0,q_w}$, which belongs to
the trivial $SU(2)_L\times SU(2)_I$ representation
and has even parity because $n_z=0$.
Hence, the tensor product state of 
$\ket{\ell=0,q_w}$ with $|n_1^{\Psi}=1\rangle$ has
$SU(2)_L\times SU(2)_R\times SU(2)_I$ spin given by
\begin{align}
  (1,1/2)^{1/2}_-\otimes(0,0)^0_+=(1,1/2)^{1/2}_- \ ,
\end{align}
where the superscripts represent the isospin.
The tensor product state of $\ket{\ell=2,q_w}$ with $|n_1^{\Psi}=1\rangle$
decomposes under $SU(2)_L\times SU(2)_R\times SU(2)_I$ as
\begin{align}
  (1,1/2)^{1/2}_-\otimes(1,0)^1_+=
\left[(2,1/2)\oplus(1,1/2)\oplus(0,1/2)\right]^{3/2}_-
\oplus 
\left[(2,1/2)\oplus(1,1/2)\oplus(0,1/2)\right]^{1/2}_-
 \ .\nn
\end{align}
It is straightforward to decompose all these states in terms
of $SU(2)_J\subset SU(2)_L\times SU(2)_R$.
The results are summarized in Table \ref{cN1/2}.
\begin{table}[h]
\begin{center}
\begin{tabular}{c||c|c}
product states & $SU(2)_L\times SU(2)_R\times SU(2)_I$
& $SU(2)_J\times SU(2)_I$
\\ \hline\hline
$|\ell=0,q_w=N_c-1\rangle\otimes|n_1^\Psi=1\rangle$  
& $(1,1/2)^{1/2}_-$ 
& $(3/2)^{1/2}_-\oplus(1/2)^{1/2}_-$
\\ \hline
$|\ell=2,q_w=N_c-1\rangle\otimes|n_1^\Psi=1\rangle$  
& $\left[(2,1/2)\oplus(1,1/2)\oplus(0,1/2)\right]^{3/2}_-$ 
& $[(5/2)\oplus 2(3/2)\oplus 2(1/2)]^{3/2}_-$
\\
& $\oplus\left[(2,1/2)\oplus(1,1/2)\oplus(0,1/2)\right]^{1/2}_-$ 
& $\oplus[(5/2)\oplus 2(3/2)\oplus 2(1/2)]^{1/2}_-$
\end{tabular}
\parbox{15cm}{
\caption{Excited baryon states for $\cN=1/2$. }
\label{cN1/2}}
\end{center}
\end{table}
Note that $(3/2)^{1/2}_-$ appearing in the first row is identified
with N(1520) in the previous section.

We next turn to discussing the $\cN=1$ states.
This is possible only when $(\cN_{84},\cN_{44})=(1,0)$
or $(0,1)$. The first condition is further divided into two cases:
(i) $(n_1^\Psi,n_1^{\ol\Psi})=(2,0),(1,1),(0,2)$ and
(ii) $(n_j^\Psi,n_j^{\ol\Psi})=(1,0),(0,1)$ for $j=2,3,4$.
Note that the 8-4 massive states with $j=2,3,4$ are given by the three
states with $N_{84}=1$ listed in Table \ref{table:95summary}.
Again, we focus on the lightest states in each case, implying that
we pick up only $(n_1^\Psi,n_1^{\ol\Psi})=(2,0)$
and $(n_j^\Psi,n_j^{\ol\Psi})=(1,0)$ with $j=2,3,4$.
The second condition $(\cN_{84},\cN_{44})=(0,1)$ is solved by
$n_k^{\Phi}=1$ with $k=1,2,3,4$
with the rest of the excitation numbers set to zero. 
Note that the 4-4 string states 
labeled by $k=1,2,3,4$ are given by the excited states with $N_{44}=1$
shown in Table \ref{table:55summary}.
As no excitation is made by any 8-4 string mode, this case gives
$q_w=N_c$.

Let us now work out the baryon states for the above three cases.
{}For the first case, the state in the massive sector is given by
$|n_1^\Psi=2\rangle$, which transforms under 
$SU(2)_L\times SU(2)_R\times SU(2)_I$ as
\begin{align}
  \left[(1,1/2)_-^{1/2}\otimes (1,1/2)_-^{1/2}\right]_{\rm symmetrized}
=\left[
(0,1)^1\oplus
(0,0)^0\oplus
(2,1)^1\oplus
(2,0)^0\oplus
(1,1)^0\oplus
(1,0)^1
\right]_+ \ .
\label{nn20}
\end{align}
The massless sector for this case is characterized by $q_w=N_c-2={\rm odd}$.
We are thus allowed to set $\ell=1$ as the lightest state, whose
$SU(2)_L\times SU(2)_R\times SU(2)_I$ spin is given by $(1/2,0)_+^{1/2}$.
By taking the tensor product of this state $\ket{\ell=1,q_w=N_c-2}$ with 
$|n_1^\Psi=2\rangle$,
we find the following baryon states:
\begin{align}
&  [
(1/2,0)\oplus
(1/2,1)\oplus
(3/2,0)\oplus
(3/2,1)\oplus
(5/2,1)
]_+^{3/2}
\nn\\
&\oplus
  [
2(1/2,0)\oplus
2(1/2,1)\oplus
2(3/2,0)\oplus
2(3/2,1)\oplus
(5/2,0)\oplus
(5/2,1)
]_+^{1/2}\ .
\end{align}

{}For the second case, we take the massive sector state
to be
$|n_j^\Psi=1\rangle$ with $j=2,3,4$. 
This corresponds to $q_w=N_c-1={\rm even}$.
We can take $\ell=0,2,4\cdots$.
The state $\ket{\ell=0,q_w}$ has a trivial spin so that the 
tensor product of this state with 
$|n_j^\Psi=1\rangle$ has the same spin as
$|n_j^\Psi=1\rangle$.
The massless sector with $\ell=2$ has $SU(2)_L\times SU(2)_R\times SU(2)_I$
spin given by $(1,0)^1$. The tensor product of this state
with $|n_j^\Psi=1\rangle$ 
is easy to evaluate for each $j=2,3,4$.

{}Finally, the massive sector for the third case is
characterized by the four states $|n_k^\Phi=1\rangle$
with $k=1,2,3,4$.
As noted before, this corresponds to $q_w=N_c={\rm odd}$ so that
odd $\ell$ is allowed.
We pick up $\ell=1$, which is expected to give the lightest
state among those with odd $\ell$, and take its tensor product
with $|n^\Phi_k=1\rangle$.
Note that any 4-4 string state has a vanishing isospin.
The same computation is easy to perform for the
next-lightest state with $\ell=3$.

All the results are summarized in Table \ref{cN1}.
\begin{table}[h]
\begin{center}
\begin{tabular}{c||c}
product states & $SU(2)_L\times SU(2)_R\times SU(2)_I$
\\ \hline\hline
$|\ell=1,q_w=N_c-2\rangle\otimes|n_1^\Psi=2\rangle$  
& 
$
[
(1/2,0)\oplus
(1/2,1)\oplus
(3/2,0)\oplus
(3/2,1)\oplus
(5/2,1)
]_+^{3/2}
$
\\
&  
$
\oplus[
2(1/2,0)\oplus
2(1/2,1)\oplus
2(3/2,0)\oplus
2(3/2,1)\oplus
(5/2,0)\oplus
(5/2,1)
]_+^{1/2}
$
\\ \hline
$|\ell=0,q_w=N_c-1\rangle\otimes|n_2^\Psi=1\rangle$  
& 
$
(1/2,0)^{1/2}_+
$
\\ \hline
$|\ell=0,q_w=N_c-1\rangle\otimes|n_3^\Psi=1\rangle$  
& 
$
(1/2,1)_+^{1/2}
$
\\ \hline
$|\ell=0,q_w=N_c-1\rangle\otimes|n_4^\Psi=1\rangle$  
& 
$
(3/2,1)_+^{1/2}
$
\\ \hline
$|\ell=2,q_w=N_c-1\rangle\otimes|n_2^\Psi=1\rangle$  
& 
$
[(3/2,0)\oplus(1/2,0)]^{3/2}_+
\oplus
[(3/2,0)\oplus(1/2,0)]^{1/2}_+
$
\\ \hline
$|\ell=2,q_w=N_c-1\rangle\otimes|n_3^\Psi=1\rangle$  
& 
$
[(3/2,1)\oplus(1/2,1)]^{3/2}_+
\oplus
[(3/2,1)\oplus(1/2,1)]^{1/2}_+
$
\\ \hline
$|\ell=2,q_w=N_c-1\rangle\otimes|n_4^\Psi=1\rangle$  
& 
$
[(5/2,1)\oplus(3/2,1)\oplus(1/2,1)]^{3/2}_+
\oplus
[(5/2,1)\oplus(3/2,1)\oplus(1/2,1)]^{1/2}_+
$
\\ \hline
$|\ell=1,q_w=N_c\rangle\otimes|n_1^\Phi=1\rangle$  
& 
$
(1/2,0)^{1/2}_+
$
\\ \hline
$|\ell=1,q_w=N_c\rangle\otimes|n_2^\Phi=1\rangle$  
& 
$
(1/2,0)^{1/2}_+
$
\\ \hline
$|\ell=1,q_w=N_c\rangle\otimes|n_3^\Phi=1\rangle$  
& 
$
\left[ (1,1/2)\oplus (0,1/2)\right]^{1/2}_-
$
\\ \hline
$|\ell=1,q_w=N_c\rangle\otimes|n_4^\Phi=1\rangle$  
& 
$
[(3/2,1)\oplus(1/2,1)]^{1/2}_+
$
\\ \hline
$|\ell=3,q_w=N_c\rangle\otimes|n_1^\Phi=1\rangle$  
& 
$
(3/2,0)^{3/2}_+
$
\\ \hline
$|\ell=3,q_w=N_c\rangle\otimes|n_2^\Phi=1\rangle$  
& 
$
(3/2,0)^{3/2}_+
$
\\ \hline
$|\ell=3,q_w=N_c\rangle\otimes|n_3^\Phi=1\rangle$  
& 
$
\left[ 
(2,1/2)
\oplus (1,1/2)
\right]^{3/2}_-
$
\\ \hline
$|\ell=3,q_w=N_c\rangle\otimes|n_4^\Phi=1\rangle$  
& 
$
[(5/2,1)
\oplus(3/2,1)
\oplus(1/2,1)
]^{3/2}_+
$
\\ \hline
\end{tabular}
\parbox{15cm}{
\caption{Excited baryon states for $\cN=1$. }
\label{cN1}}
\end{center}
\end{table}
Decomposing these states in terms of $SU(2)_J\subset SU(2)_L\times SU(2)_R$
is straightforward. 

Now we discuss possible identifications of the
states listed in Tables \ref{cN1/2} and \ref{cN1}
with the baryons found in experiments.
Because we have not been able to derive the $\ell$ dependence
in the baryon mass formula (\ref{mass}), we have to rely on
some qualitative arguments. Our guiding principles are as follows.
First, we expect that the states with the same $\ell$, $q_w$ and $\cN$
are nearly degenerate.
Second, for a given $(\ell,q_w)$, the states with $\cN=1$ are
heavier than those with $\cN=1/2$. Third, for a given
$\cN$, the mass is an increasing function of both $\ell$
and $q_w$ except for the state with $n_1^\Psi=2$ listed in the first row
of Table \ref{cN1}, which is expected to be heavier than the others
according to the baryon mass formula (\ref{mass}).
\footnote{
Here, we have assumed that $M_0^*|_{q_w=N_c}-M_0^*|_{q_w=N_c-2}$ is
smaller than $(\sqrt{2}-1)/\alpha'$, which can be justified for
large $\lambda$.}

The predictions for the low-lying excited baryons with $I=1/2$ are
summarized in Table \ref{I1/2}, whose data are taken from
Tables \ref{cN1/2} and \ref{cN1}.
\begin{table}[H]
%\vspace{0.1cm}
\begin{center}
  \begin{tabular}{c|l||c}
level & states & $SU(2)_J\times SU(2)_I$
\\ \hline\hline
$\cN=1/2$
&
$|\ell=0,q_w=N_c-1\rangle\otimes|n_1^\Psi=1\rangle$
&
$[{\color{blue} (3/2)}\oplus (1/2)]_-^{1/2}$
\\ \cline{2-3}
&
$|\ell=2,q_w=N_c-1\rangle\otimes|n_1^\Psi=1\rangle$
&
$[(5/2)\oplus 2(3/2)\oplus 2(1/2)]_-^{1/2}$
\\ \hline
$\cN=1$
&
$|\ell=0,q_w=N_c-1\rangle\otimes|n_{2,3,4}^\Psi=1\rangle$
&
$[{\color{blue}(5/2)}\oplus 2(3/2)\oplus 3(1/2)]_+^{1/2}$
\\ \cline{2-3}
&
$|\ell=1,q_w=N_c\rangle\otimes|n_{1,2,3,4}^\Phi=1\rangle$
&
$
[(5/2)\oplus 2(3/2)\oplus 4(1/2)]_+^{1/2}
\oplus
[(3/2)\oplus 2(1/2)]_-^{1/2}
$
\\ \cline{2-3}
&
$|\ell=2,q_w=N_c-1\rangle\otimes|n_{2,3,4}^\Psi=1\rangle$
&
$
[(7/2)\oplus 3(5/2)\oplus
6(3/2)\oplus 5(1/2)]_+^{1/2}
$
\\ \hline
  \end{tabular}
\parbox{15cm}{
\caption{Low-lying excited baryons with $I=1/2$. Here we have omitted
the states with $n_1^\Psi=2$ in Table \ref{cN1}.
The blue-colored states are identified with excited baryons lying
on the nucleon Regge trajectory in section \ref{Regge}.
}\label{I1/2}}
\end{center}
\end{table}
We will not attempt to relate the states 
with $J=1/2$ in this table to those in the baryon summary
table \cite{pdg}, here, 
because these might be regarded as excited states
with nonvanishing $n_\rho$, $n_z$ and $n_\beta$ without
excitations in the massive sector.\footnote{
Some such states were already discussed in \cite{HSSY}.}
Note that the states with $(3/2)_-^{1/2}$ and $(5/2)_+^{1/2}$
in the first and third rows of Table \ref{I1/2} are identified with
N(1520) and N(1680), respectively, in section \ref{Regge}.
The $(5/2)_-^{1/2}$ state at $\cN=1/2$ is expected to be the
lightest state with this quantum number and hence it may be
identified with N(1675), which is the lightest baryon
with the same quantum number listed in the baryon summary table.
Then, the $(3/2)_-^{1/2}$ states in the second row
are expected to have mass nearly equal to N(1675).
A natural candidate for one of them is N(1700).\footnote{
There are other possibilities for this identification. For example,
$\ket{\ell=0,n_\rho=1,q_w=N_c-1}\otimes\ket{n_1^\Psi=1}$
and
$\ket{\ell=3,n_z=1,n_\beta=1,q_w=N_c}$ also have $(3/2)_-^{1/2}$
components that could be identified with N(1700).
}

{}As for the $\cN=1$ states, we find that 
the $(3/2)_+^{1/2}$ states in the third row of Table \ref{I1/2} are
expected to have mass nearly equal to N(1680).
A natural candidate for one of them is N(1720).\footnote{
As in the case of N(1700),
$\ket{\ell=0,n_z=1,q_w=N_c-1}\otimes\ket{n_1^\Psi=1}$
and
$\ket{\ell=3,n_\rho=1,n_\beta=1,q_w=N_c}$ also have $(3/2)_+^{1/2}$
components that could be identified with N(1720).
}
Since the fourth row has larger values of $\ell$ and $q_w$
compared with the third row,
the $(5/2)_+^{1/2}$ state in the fourth row is expected to be
heavier than N(1680) and N(1720).
A natural candidate for it is N(1860), though this state
has not been established in experiments. If this is the case,
the $(3/2)_\pm^{1/2}$ states in the fourth row are expected
to be nearly degenerate with N(1860). These states could be identified
with N(1900) and N(1875).
The baryon states in the fifth row contain a state with
$(7/2)_+^{1/2}$. The only baryon with this quantum number
listed in the baryon summary table is N(1990),
though this is not considered to be established.
Then, the $(5/2)_+^{1/2}$ and $(3/2)_+^{1/2}$ states
in the fifth row of Table \ref{I1/2}
could be identified with N(2000) and N(2040), respectively,
which are again poorly established in experiments.

Unfortunately, the identification we have made is not a clear one-to-one
correspondence. There is more than one
candidate state in the model for many of the baryons listed in the
baryon summary table. In particular, 
the degeneracy of the states in Table \ref{I1/2} does not match
the experimental data perfectly.
Furthermore, as mentioned in the footnotes, some of the baryons
may be identified with the states that are not listed
in Table \ref{I1/2}. 
This lack of the one-to-one correspondence
could in part be because all the excited baryons we consider
are unstable resonances (for finite $N_c$), and many of them, in
particular the heavier ones, are probably not easy to identify in
experiments.
{}Furthermore, some of the states in Tables \ref{I1/2} and $\ref{I3/2}$
could be artifacts of the model. Although, as discussed in section
\ref{sec:baryonstates}, we have imposed invariance with respect to
the $SO(5)$ symmetry and $\tau$-parity to get rid of artifacts, we
are not able to show that this is sufficient to exclude all of them.
It is expected that incorporation of full $1/\lambda$ corrections 
into the baryon mass formula makes the artifacts of the model
infinitely heavy in the $\Mkk\rightarrow\infty$ ($\lambda\rightarrow 0$)
limit with $\Lambda_{\rm QCD}$ kept fixed. However, the extrapolation
to the small $\lambda$ regime is a notoriously difficult problem
in the holographic description, because we have to deal with
all the stringy corrections in a highly curved spacetime.
A similar observation was also made in \cite{ISS}.
We leave as an open problem the study of
a dictionary between the theoretical predictions and
the experimental data in more detail.

We also examine the mass spectrum of $\Delta$ baryons with
isospin $I=3/2$. The theoretical predictions for this case are summarized
in Table \ref{I3/2},
whose data is taken from Tables \ref{cN1/2} and \ref{cN1}.
\begin{table}[H]
%\vspace{0.1cm}
\begin{center}
  \begin{tabular}{c|l||c}
level & states & $SU(2)_J\times SU(2)_I$
\\ \hline\hline
$\cN=1/2$
&
$|\ell=2,q_w=N_c-1\rangle\otimes|n_1^\Psi=1\rangle$
&
$[(5/2)\oplus 2(3/2)\oplus 2(1/2)]_-^{3/2}$
\\ \hline
$\cN=1$
&
$|\ell=2,q_w=N_c-1\rangle\otimes|n_{2,3,4}^\Psi=1\rangle$
&
$[(7/2)\oplus 3(5/2)
\oplus 6(3/2)\oplus 5(1/2)]_+^{3/2}$
\\ \hline
  \end{tabular}
\parbox{15cm}{\caption{Low-lying excited baryons with $I=3/2$.
In this table, we have omitted
the states with $n_1^\Phi=2$ and $\ell=3$ in Table \ref{cN1}.
}\label{I3/2}}
\end{center}
\end{table}
It is natural to identify the $(5/2)_-^{3/2}$ and $(7/2)_+^{3/2}$ states
in the first and second rows of Table \ref{I3/2} with the lightest
$\Delta$ baryons having the same quantum numbers
listed in the baryon summary table, which are
$\Delta(1930)$ and $\Delta(1950)$, respectively.
This suggests that the $(5/2)_+^{3/2}$ states in the second
row of Table \ref{cN1} are nearly degenerate with
$\Delta(1950)$. A good candidate to be identified with
one of these states is $\Delta(1905)$.
However, this identification is problematic:
although our formula (\ref{mass}) suggests that the $\cN=1$ states are
significantly heavier than the $\cN=1/2$ states,
$\Delta(1930)$ and $\Delta(1950)$ are nearly degenerate
and $\Delta(1905)$ is even lighter than $\Delta(1930)$.

%%%%%%%%%%%%%%%%%%%%%%%%%%%%%%%%%%%%%%%%%%%%%%%%%%%%%%%%%%%%%%%%%%%%
%%%%%%%%%%%%%%%%%%%%%%%%%%%%%%%%%%%%%%%%%%%%%%%%%%%%%%%%%%%%%%%%%%%%
\section{Conclusions}

We have discussed stringy excited baryons using
the holographic dual of QCD on the basis of an intersecting
D4/D8-brane system.
A key step to this end is to work on the whole system of
a baryon vertex without describing it by a topological
soliton on an effective 5 dimensional gauge theory.
We formulated this system as a many-body quantum mechanics
that is composed of the ADHM-type matrix model of Hashimoto-Iizuka-Yi
\cite{HIY} and an infinite number of open string massive modes.
This is done by relying on an approximation that is
valid in the large $N_c$ and $\lambda$ regime.
The resultant quantum mechanics provides us with a powerful
framework for making a systematic analysis of excited baryons
including those with $I\ne J$ that are difficult to obtain in the
soliton picture.

By construction, it would be too ambitious for the theoretical
predictions from the present model to match the experimental data 
to good accuracy. Interestingly, we have seen that the present
model reproduces a qualitative feature of the nucleon Regge trajectory.
It has been argued that the stringy excited baryons to be identified
with the excited nucleons are interpreted as a rotating open string
with a massive end point. Such a picture of baryon Regge trajectories
has been studied extensively in the
literature. \cite{Sharov:2013tga,cobi} It is worth 
emphasizing that the massive end point in this model is due to
a D4$_{\rm BV}$, having a mass of $\cO(N_c)$.
The Regge trajectory formula (\ref{JE}) that we proposed in this paper
is not given by a simple, linear relation between the spin and the mass
squared because of the heavy end point.

We conclude this paper by making some comments about
future directions. First, it is important to
improve the theoretical accuracy of the model by incorporating
the interacting terms in $L_{\rm int}$ that have been neglected 
for technical difficulties. It would be almost impossible to
fix the mass terms of the mass fields $\Psi_j$ and $\Phi_k$ 
precisely, because infinitely many higher-order terms could
contribute to a single mass term, as discussed in section
\ref{subsec:Lint}.
Instead, what may be performed immediately is to take into
account the effects of the mixing terms like $\Psi_j\Psi_j$
in the baryon mass formula.
With these mixing terms, $q_j$ is not a conserved charge any more
so that an exact diagonalization of $H_m$ in a manner
consistent with the Gaussian constraint appears highly involved.
It would be interesting to compute the perturbative effect of the mixing
terms into the mass formula.

One of the unsatisfactory points is that the values of the parameters
$\gamma$ and $v$ in the potential (\ref{V0}) are not determined from
first principles. Though it is possible to adjust them to fit the
results in the soliton picture as in \cite{HSSY}, a derivation within
our framework is desired to make sure that all the parameters can be
fixed, in principle, without any ambiguities.
Compared with the soliton picture, the origin of the potential
(\ref{V0}) is expected to be due to the energy contribution from the
$U(N_f)$ gauge field on the flavor D8-branes in the presence of a baryon
vertex. It would be interesting to examine this in more detail.

{}Finally, it would be of great interest to apply the results in
this paper to a more complicated system made out of
multiple baryon and anti-baryon vertices.
A typical example is given by a stringy realization
of tetraquarks. It would be nice to try to formulate a holographic model 
for tetraquarks following this paper and
compare the theoretical predictions with experiments.

%\afterpage{\clearpage}
%\newpage

%%%%%%%%%%%%%%%%%%%%%%%%%%%%%%%%%%%%%%%%%%%%%%%%%%%%%%%%%%%%%%%%%%%%
%%%%%%%%%%%%%%%%%%%%%%%%%%%%%%%%%%%%%%%%%%%%%%%%%%%%%%%%%%%%%%%%%%%%
%
\section*{Acknowledgments}
We would like to thank K. Hashimoto, S. Hirano and J. Sonnenschein
for useful discussions.
The work of SS was supported by JSPS KAKENHI
(Grant-in-Aid for Scientific Research (C)) grant number JP16K05324
and
(Grant-in-Aid for Scientific Research (B)) grant number JP19H01897.

%%
%%%%%%%%%%%%%%%%%%%%%%%%%%%%%%%%%%%%%%%%%%%%%%%%%%%%%%%%%%%%%%%%%%%%
%%%%%%%%%%%%%%%%%%%%%%%%%%%%%%%%%%%%%%%%%%%%%%%%%%%%%%%%%%%%%%%%%%%%
%
\appendix
\section{$SO(4)\simeq SU(2)_I\times SU(2)_J$}
\label{SO4}

The generators of the Lie algebra of
$SO(4)\simeq (SU(2)_I\times SU(2)_J)/\bZ_2$ can be chosen as
\begin{eqnarray}
 i\Sigma^1&=&i\sigma_2\otimes\sigma_1
=\mat{,\sigma_1,-\sigma_1,}\ ,\nn\\
 i\Sigma^2&=&-i\sigma_2\otimes\sigma_3
=\mat{,-\sigma_3,\sigma_3,}\ ,\nn\\
 i\Sigma^3&=&i{\bf 1}_2\otimes\sigma_2
=\mat{i\sigma_2,,,i\sigma_2}\ ,
\end{eqnarray}
\begin{eqnarray}
 i\wt\Sigma^1&=&-i\sigma_1\otimes\sigma_2
=\mat{,-i\sigma_2,-i\sigma_2,}\ ,\nn\\
 i\wt\Sigma^2&=&-i\sigma_2\otimes{\bf 1}_2
=\mat{,-{\bf 1}_2,{\bf 1}_2,}\ ,\nn\\
 i\wt\Sigma^3&=&i\sigma_3\otimes\sigma_2
=\mat{i\sigma_2,,,-i\sigma_2}\ .
\end{eqnarray}

$\{\Sigma^a\}$ and $\{\wt\Sigma^a\}$ satisfy
the same algebra as the Pauli matrices
\begin{eqnarray}
 \Sigma^a\Sigma^b=\delta^{ab}+i\epsilon^{abc}\Sigma^c\ ,~~~
\wt\Sigma^a\wt\Sigma^b=\delta^{ab}+i\epsilon^{abc}\wt\Sigma^c\ ,
\end{eqnarray}
and they commute with each other
\begin{eqnarray}
 \Sigma^a\wt\Sigma^b&=&\wt\Sigma^b\Sigma^a\ .
\end{eqnarray}
$\{i\Sigma^a\}_{a=1,2,3}$ and
$\{i\wt\Sigma^a\}_{a=1,2,3}$ are the generators of
$SU(2)_I$ and $SU(2)_J$, respectively.

%
%%%%%%%%%%%%%%%%%%%%%%%%%%%%%%%%%%%%%%%%%%%%%%%%%%%%%%%%%%%%%%%%%%%
%%%%%%%%%%%%%%%%%%%%%%%%%%%%%%%%%%%%%%%%%%%%%%%%%%%%%%%%%%%%%%%%%%%
%

%
%
\end{document}